\documentclass[twocolumn,a4paper,11pt]{article}
\usepackage{blindtext}
\usepackage[utf8]{inputenc}
\usepackage{times}
\usepackage{microtype}

\usepackage{lscape}

\usepackage{epstopdf}

\usepackage{natbib}
\bibpunct{(}{)}{;}{a}{,}{,}
\usepackage{subfigure}
\usepackage{multirow}
\usepackage{float}
\usepackage{soul}
\usepackage{xcolor}
\graphicspath{{./images/}}

\usepackage{amssymb}
\usepackage{amsmath}	
\usepackage{mathtools}
\usepackage{amsthm}
\usepackage{bm}
\usepackage{url}
\usepackage{parskip}

\usepackage[titletoc]{appendix}
\usepackage{booktabs}
\usepackage{boldline}
\usepackage{colortbl}

\providecommand{\keywords}[1]
{
  \small	
  \textbf{\textit{Keywords---}} #1
}

\usepackage{listings}
\usepackage{caption}
\begin{document}

\title{A Bayesian Multilevel Random-Effects Model for Estimating Noise in Image Sensors}

\author{Gabriel Riutort-Mayol$^{1*}$, Virgilio G\'omez-Rubio$^{2}$, \'Angel Marqu\'es-Mateu$^{1}$, \\ Jos\'e Luis Lerma$^{1}$, Antonio L\'opez-Qu\'ilez$^{3}$}

\date{ \small
$^1$ Department of Cartographic Engineering, Geodesy, and Photogrammetry, Universitat Polit\`ecnica de Val\`encia, Spain 
\break
$^2$ Department of Mathematics, Universidad de Castilla-La Mancha, Albacete, Spain
\break
$^3$ Department of Statistics and Operations Research, Universitat de Val\`encia, Burjassot, Spain
\break
$^*$ Corresponding author, Email: gabriuma@gmail.com
}

\maketitle

\makeatletter{\renewcommand*{\@makefnmark}{}
\footnotetext{\textit{\small This paper is a postprint of a paper submitted to and accepted for publication in IET Image Processing and is subject to Institution of Engineering and Technology Copyright. The copy of record is available at the IET Digital Library.}}\makeatother}

\begin{abstract}
Sensor noise sources cause differences in the signal recorded across pixels in a single image and across multiple images. This paper presents a Bayesian approach to decomposing and characterizing the sensor noise sources involved in imaging with digital cameras. A 
Bayesian probabilistic model based on the (theoretical) model for noise 
sources in image sensing is fitted to a set of 
a time-series of images with different reflectance and wavelengths 
under controlled lighting conditions. The image sensing model is a 
complex model, with several interacting components dependent on 
reflectance and wavelength. The properties of the Bayesian approach of defining conditional dependencies among parameters in a fully probabilistic model, propagating all sources of uncertainty in inference, makes the Bayesian modeling framework more attractive and powerful than classical methods for approaching the 
image sensing model. A feasible correspondence of noise parameters to 
their expected theoretical behaviors and well calibrated posterior 
predictive distributions with a small root mean square error for model predictions have been achieved in this study, thus showing that the proposed model accurately approximates the image sensing model. The Bayesian approach could be extended to formulate further components aimed at identifying even more specific parameters of the imaging process.
\end{abstract}


\section{Introduction}\label{ch3:intro}

Some industrial applications require calibrated image sensors. An accurate procedure to model the image sensor noise is useful for optimizing sensor design as well as for finding out how much uncertainty, in their different sources, is present in an image \citep{european2010emva,dierks2004sensitivity,kuroda2014essential}. These noise sources are reflected on differences in the signal recorded accross pixels in a single image and accross multiple images coming from different shoots (image captures), and the variances of these noises are dependent on the level of reflectance imaged.

Image sensing and its different noise components are well documented in the literature, where their manifestations and relationships are well defined.
\cite{de2011noise,reibel2003ccd,aguerrebere2013study} and \cite{dierks2004sensitivity} are excellent introductory references for understanding and extracting a model for image sensing. Based on those references, in Section~\ref{ch3:sensingmodel} a conceptual model for all the components involved in image sensing data is described and formulated. This conceptual model will be used as the data-generating model for the proposed model in Section \ref{ch3:multilevel_model}. 

A number of excellent references provide models for the sensing process that additionally include the description of other stages in the practice of imaging applications, such as the optical system of the camera~\citep{campos2000radiometric}, image processing parameters~\citep{tsin2001statistical} and reflectance and illumination variations~\citep{healey1994radiometric}. There are many other references, from different disciplines such as computer vision, photometry, physics, and electronics, dealing with the definition of all the parameters involved in the sensing process~\citep{grant2005characterization,chi2011noise,han2011research,granados2010optimal}. The fact is that their interpretations agree on the process of image sensing. Other related relevant research applications such as image noise removal can be found in \cite{ghita2012adaptive,shukla2018denoising,zhang2017image}.

In order to estimate the effects and contributions of the sensor noise parameters on the output image, the data-generating model of the process is fitted to a set of observed image data. For this purpose, classical iterative optimization algorithms based on maximum likelihood and point estimates of the parameters \citep{healey1994radiometric,tsin2001statistical,dong2018effective}, and the well-known Photon transfer method~\citep{janesick1985ccd}, are the current methods used for fitting and inferring those noise components. In particular, the Photon transfer method is the technique used to characterize the noise parameters in the ISO \citep{international2003iso} and EMVA \citep{european2010emva} standards for image noise measurements. The Photon transfer method \citep{reibel2003ccd,dierks2004sensitivity,de2011noise} is based on averages across the spatial-arranged matrix of pixels as well as on averages across different image exposures, of dark and illuminated images. A possible weakness of the Photon transfer method is that it is based on nested independent and point estimate computations, which induces probably high error propagation, apart from the need to make assumptions such as normality and independence between some parameters (i.e. photo response non-uniformity and photon noise).

The present study is focused on Bayesian modeling and inference of those sensor noise components. In the Bayesian context \citep{jaynes2003probability,bernardo2009bayesian,Gelman_2013}, all inference is based on the (multivariate) posterior distribution of model parameters and hyperparameters. Computing the posterior distribution is often difficult and, for this reason, different computational approaches can be used. Markov chain Monte Carlo (MCMC) \citep{brooks_2011} are sampling methods that provide samples of the joint posterior distribution of parameter $\theta$ given data $y$, $\pi(\theta|y)$. These samples can be used to make inference and assess the significance of the different parameters in the model. The Bayesian approach quantifies uncertainty in inferences through probability distributions. In addition, interactions between different terms can be easily explored by means of their joint posterior distributions.

Bayes estimates have many advantages compared to point estimates of classical methods \citep{raiko2006bayesian,bishop2006pattern,browne2006comparison,Gelman_2013}. Point estimates use a single representative value to summarize the whole posterior distribution. Maximum likelihood finds the point estimate of the parameters that maximize the observed data, $\pi(y|\theta)$. In contrast, Bayesian inference estimates the whole posterior join distribution of the parameters given the data, $\pi(\theta|y)$. The problem of overfitting is mostly related to point estimates. The use of a point estimate is to approximate integrals so it should be sensitive to the probability mass rather than to the probability density. Maximum likelihood estimates are attracted to high but sometimes narrow peaks and, unfortunately, this effect becomes stronger when the dimensionality increases \citep{raiko2006bayesian,bishop2006pattern}.

The real strength of the Bayesian approach comes from the possibility of constructing hierarchical models, also known as multilevel models, which may define models with complex structures by defining conditional dependencies among quantities \citep{brown2014applied,gelman2006data,coley2017bayesian,dai2017predicting}, while propagating all sources of uncertainty for inference. Bayesian hierarchical models naturally lead to more reliable inferences and better real-world answers \citep{Gelman_2013,browne2006comparison}. All these advantages motivate the present work and the use of a Bayesian approach for the study of image sensing noise components.

In the present work, a novel Bayesian approach in the field of image sensor noise characterization is performed. A probabilistic model based on the data-generating model is fitted to a set of a time-series of images with different reflectance and wavelengths under uniform illumination conditions. The data-generating model adds several and interacting random components dependent on reflectance and wavelength, so a Bayesian hierarchical model with conditional dependencies among model parameters, i.e. a multilevel random-effects model, is a suitable model and the model proposed to approximate the data-generating model. The unknown parameters in the probabilistic model, which are the parameters of the noise components of the process, are learned through sampling methods based on MCMC. The flexibility, accuracy, and intuitiveness of the Bayesian framework for modeling and calibrating the sensor noise components is worth mentioning. The results show a reliable and flexible modeling, able to naturally and accurately propagate uncertainties of noise parameters. 

The rest of the paper is structured as follows. Section~\ref{ch3:sensingmodel} reviews all the noise components, their manifestations and relationships, formulating the data-generating model (theoretical model) for image sensing. Section \ref{ch3:experiment} describes the available experimental data. Section \ref{ch3:proposed} focuses on the modeling and inference formulation of the proposed Bayesian multilevel random-effects model. Section \ref{ch3:results} analyzes the results of fitting the proposed model to the experimental data. Section \ref{ch3:modelchecking} describes the procedures used for model checking and assessment. Section \ref{ch3:discussion} discusses the standards for image noise measurements and makes a brief qualitative comparison of the standards with the proposed statistical modeling. Finally, Section \ref{ch3:conclu} draws some conclusions.

\section{Image sensing model} \label{ch3:sensingmodel}

A digital image is formed once the electromagnetic energy coming from or reflected by an object is registered into an image sensor at a certain instant after shooting (image capture). 
The reflectance of an object  
is usually considered as a continuous factor between 0 and 1, where zero represents null reflectance and one total reflectance~\citep{pratt2007digital}.  
An image sensor is composed of many individual sensing elements (pixels) arranged in a regular matrix that register incoming light at a certain instant or shoot.

Basically, photons of energy emitted from and reflected by the object are captured by a single pixel. Each one of the photons inside the pixel has a probability, called quantum efficiency, to create a free electron. Then, from the incoming photons, a number of electrons are created inside the pixel. Finally, the electrons, after being converted into a voltage, are amplified and digitized into an output digital number, also known as gray level \citep{dierks2004sensitivity,tsin2001statistical,healey1994radiometric}.

Following \cite{de2011noise}, \cite{reibel2003ccd} and \cite{dierks2004sensitivity}, a simple model of the output digital numbers $y_{it}$ registered in the $i$'th pixel and at the $t$'th image shoot, as a function of the reflectance $r$ of the reflective object and the wavelength $w$ of the light, can be written as follows:
\begin{align} \label{ch3:eq:sensingmodel}
 y_{it}(r,w) =& \; K_i \hspace{-0.5mm} \cdot \hspace{-0.5mm} e_{it}(r,w) + \mu_{\scriptscriptstyle K} \hspace{-0.5mm} \cdot \hspace{-0.5mm} D_{i} + \mu_{\scriptscriptstyle K} \hspace{-0.5mm} \cdot \hspace{-0.5mm} C_{t}(r,w) 
 \nonumber \\
 & + \mu_{\scriptscriptstyle K} \hspace{-0.5mm} \cdot \hspace{-0.5mm} R_{it} + A_{it}.
\end{align}

The number of electrons $e_{it}(r,w)$ is a function of the number of photons coming into the pixel and of the probability $q(w)$ of creating a free electron from an incoming photon by the pixel sensing element. A model for the electrons is usually approximated as a Poison model
\begin{equation} \label{ch3:eq:electrons}
e_{it}(r,w) \sim Po \big(q(w) \cdot \mu_{\scriptscriptstyle p}(r,w)\big) = Po\big(\mu_{\scriptscriptstyle e}(r,w)\big),
\end{equation}

\noindent where $\mu_{\scriptscriptstyle e}(r,w)$ is the mean number of the electrons $e$ created from the incoming photons inside the pixel. $\mu_{\scriptscriptstyle p}(r,w)$ is the mean number of the incoming photons which are dependent on the reflectance $r$ and the wavelength $w$. The probability $q(w)$ also depends on the wavelength of the light. The variances $\mu_{\scriptscriptstyle p}(r,w)$ of these Poisson variables are called photon noise and represent the variances of the incoming energy in function of reflectance and wavelength. Note that the variance of $\mu_{\scriptscriptstyle e}(r,w)$ also represents the photon noise, since it is directly proportional to the mean number of photons. Moreover, it should be noted that photon noise is always present in images and is never dependent on the camera sensor.

The gain factor variable $K_{i}$ governs the process of converting electrons in a pixel into voltage, its amplification and digitalization \citep{reibel2003ccd,dierks2004sensitivity,healey1994radiometric}. There is evidence in the literature of considering $K_{i}$ contaminated with Gaussian noise (\ref{ch3:eq:PRNU}) which represents one part of the spatial noise of image sensors, commonly named photo response non-uniformity (PRNU). PRNU models the inter-pixel differences when generating electrons from the incoming photons \citep{reibel2003ccd,gow2007comprehensive,dierks2004sensitivity}, which are due to pixel pitch and other pixel characteristics \citep{dierks2004sensitivity,gow2007comprehensive}.
\begin{equation}\label{ch3:eq:PRNU}
K_{i} \sim N(\mu_{\scriptscriptstyle K},\sigma^2_{\scriptscriptstyle K})
\end{equation}

\noindent In the previous equation (\ref{ch3:eq:PRNU}), $\mu_{\scriptscriptstyle K}$ and $\sigma^2_{\scriptscriptstyle K}$ are the mean and variance of variable $K_i$. 

In addition to the electrons $e_{it}(r,w)$ generated from the incoming light energy, current noise $C_{t}(r,w)$ is an effect by which free electrons can be thermally generated during the exposure time \citep{de2011noise,reibel2003ccd,dierks2004sensitivity,gow2007comprehensive}  in the $t$'th image shoot. It is related to the temperature at a certain instant or shoot and is expected to be an effect varying only on the temporal dimension $t$, being constant across pixels~\citep{de2011noise,gow2007comprehensive} . Establishing long intervals between shoots and small exposure times, trying to maintain low temperatures in the sensor, $C_{t}(r,s)$ could be considered random and modeled as a Poisson stochastic variable \citep{international2003iso,european2010emva,marques2013statistical}. Temperature inside a pixel depends on the incoming light \citep{dierks2004sensitivity,de2011noise,gow2007comprehensive}, then current noise will be an effect dependent on reflectance $r$ and wavelength $w$. Thus, the model for current noise takes the form:
\begin{equation} \label{ch3:eq:Current}
C_{t}(r,w) \sim Po(\mu_{\scriptscriptstyle C}(r,w))
\end{equation}

\noindent where $\mu_{\scriptscriptstyle C}(r,w)$ is the mean of variable $C_t(r,w)$ as a function of reflectance and wavelength. 

Apart from the light induced electrons, dark electrons $D_{i}$ are generated in the $i$'th pixel without the presence of incident light. They are generated from dark current variations across pixels, and commonly named fixed pattern noise (FPN). This is an effect affecting the spatial dimension, being the same in all different frames or shoots \citep{dierks2004sensitivity,de2011noise,el1998modeling}. Although some cameras may have some kind of non-random spatial pattern \citep{campos2000radiometric}, for most camera sensors this spatial pattern is completely random \citep{el1998modeling} following a Poisson model:
\begin{equation} \label{ch3:eq:FPN}
D_{i} \sim Po(\mu_{\scriptscriptstyle D}),
\end{equation}

\noindent where $\mu_{\scriptscriptstyle D}$ is the mean of variable $D_i$. 

Moreover, reset noise $R_{it}$ refers to the remaining electrons in the circuitry capacitors even after being emptied in the previous exposure. It is expected to be an effect defined independently on both dimensions $i$ and $t$ and completely random, so modeled by a Poisson variable:
\begin{equation} \label{ch3:eq:Reset}
R_{it} \sim Po(\mu_{\scriptscriptstyle R}),
\end{equation}

\noindent where $\mu_{\scriptscriptstyle R}$ is the mean of variable $R_{it}$. 

The parameters $D_{i}$, $C_{t}(r,w)$ and $R_{it}$ are multiplied in equation (\ref{ch3:eq:sensingmodel}) by the mean gain parameter $\mu_{\scriptscriptstyle K}$, to encapsulate the process of converting electrons into digital numbers.

Finally, after the charge is transferred, and converted into voltage, amplified and digitized, the noise effects amplifier, $1/f$ (flicker noise \citep{han2011research}) and quantization add also some noise $A_{it}$ to the final output digital number \citep{han2011research,dierks2004sensitivity,de2011noise}. They are expected to be random and normally distributed
\begin{equation} \label{ch3:eq:Ampli}
A_{it} \sim N(\mu_{\scriptscriptstyle A},\sigma^2_{\scriptscriptstyle A}),
\end{equation}

\noindent where $\mu_{\scriptscriptstyle A}$ and $\sigma^2_{\scriptscriptstyle A}$ are the mean and variance of the variable $A_{it}$.

The variabilities of $K_i$ (PRNU), $D_i$ (FPN), $e_{it}(r,w)$ (photon noise), $C_t(r,w)$ (current noise), $R_{it}$ (reset noise), and $A_{it}$ (amplifier, 1/f and quantization noises) will be the essential parameters of an image sensor and the quantities of interest to be estimated from the model as noise parameters in this work. Photon noise is always present in an image and is never dependent on the camera sensor. The other noise parameters are dependent on the camera sensor, and so will be the parameters to compare the quality of different image sensors. The quantum efficiency is also clearly a very important parameter of quality, although it can only be estimated with the measurement, by means of a radiometer device, of the number of incoming photons into any individual pixel.

\section{Experimental data}\label{ch3:experiment}

\begin{figure}
\centering
  \includegraphics[width=7cm]{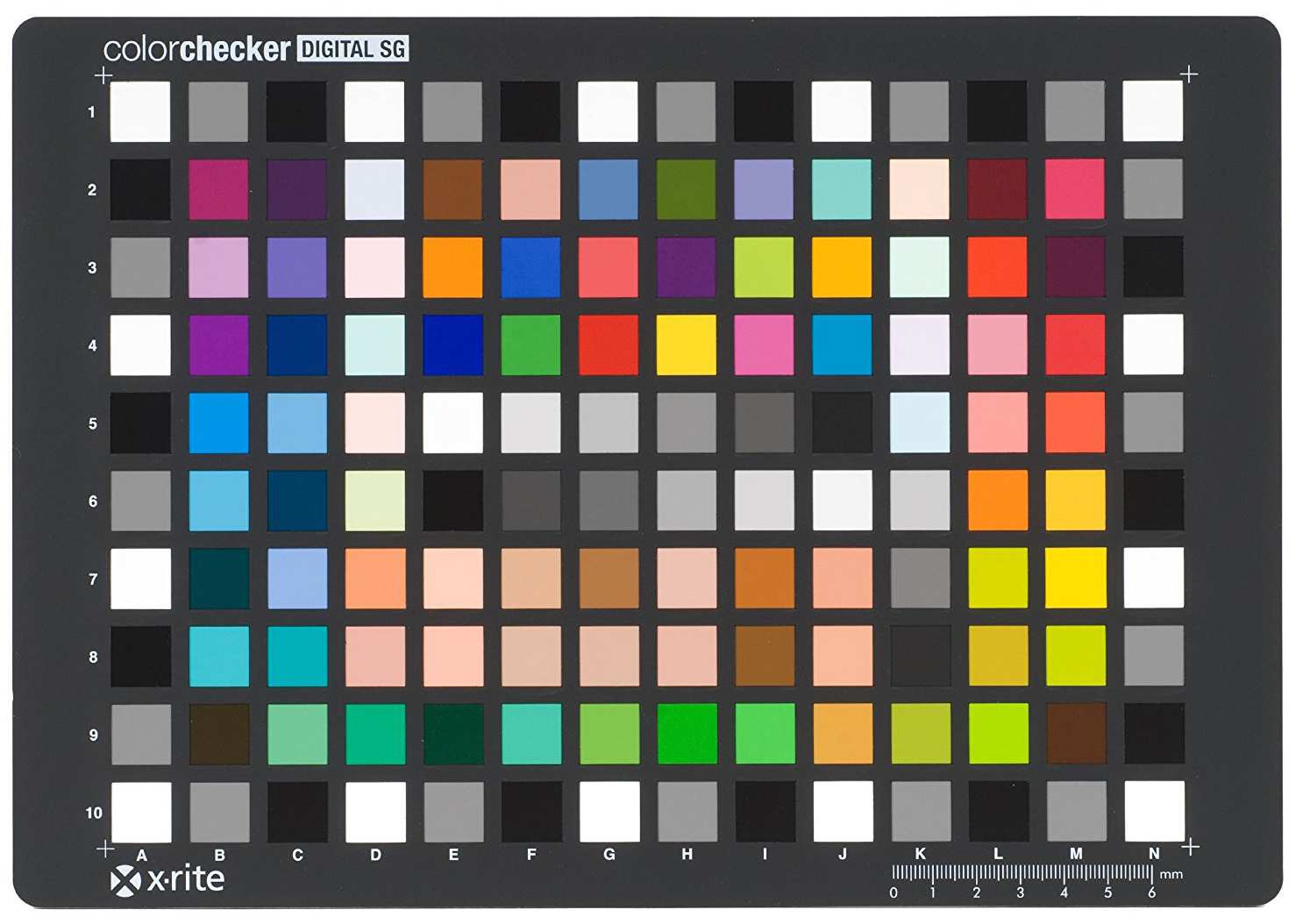}
\caption{Reflectance calibration pattern.}
\label{fig03}  
\end{figure}

The experiment consisted of time-sequential imaging of a ColorChecker by using a trichromatic image sensor camera. A ColorChecker is a reflectance calibration pattern which contains several reflectance patches, each one with constant reflectance (Figure \ref{fig03}).
A trichromatic colorimeter provides simultaneous measurements of three primary wavelength ranges (usually Red $R$, Green $G$, and Blue $B$). The result of the experiment is a time series of images with a spatially arranged matrix of pixel-values across the sensor in each image with different reflectance and wavelengths.

The experimental data was comprised of 60 images from different shoots ($t$=1,...,60). Samples of 500 random pixels ($i$=1,...,500) from 11 different reflectance patches ($r$=1,...,11) were provided for each image, resulting in 5500 pixels through the sensor, $500$ grouped pixels for each one of the 11 reflectance patches. Finally, three different wavelength ranges of the light were used ($w$=1,2,3) for each pixel.

One hundred out of this five hundred pixels within each reflectance patch were used as testing observations to make posterior model checking and validation in Section \ref{ch3:modelchecking}. Therefore, 400 pixels in each one of the reflectance patches were used to fit the model and 100 for assessing model performance.

In order to get uniform average conditions on the experiment, stable and homogeneous incident light on both dimensions, spatial and temporal, was needed. The experiment was conducted under laboratory conditions using a typical colorimetry setup following the recommendations of the \textit{Commision Internationale de l'\'Eclaraige} \citep{CIE_2004}. 

The imaging device used in the experiments was the Foveon X3\textregistered~Pro 10M CMOS sensor which has a stack of three photosensitive layers and provides true trichromatic imagery. It is considered as a high-class device that provides extremely low-noise readout and removes fixed pattern noise associated with other CMOS sensors \citep{merrill1999color}. The dynamic range of the sensor is 12 bits (0-4095 digital numbers), the total number of pixel sensors is 2268 columns x 1512 rows x 3 layers, or 3.4 million pixels per layer, and the pixel pitch of the array is 9.12 $\mu$m. This sensor also provides other interesting practical features such as low power consumption, variable pixel size, and blooming immunity.

\section{Proposed modeling and inference} \label{ch3:proposed}

\subsection{Multilevel random-effects model} \label{ch3:multilevel_model}

A multilevel random-effects model to approach the theoretical model in equation (\ref{ch3:eq:sensingmodel}) and its components is proposed. Previously, if  the Poisson variables $e_{it}(r,w)$ in (\ref{ch3:eq:sensingmodel}) are approximated as Normal variables,
\begin{equation} \label{ch3:eq:photon2}
  e_{it}(r,w) \sim N \big(\mu_{\scriptscriptstyle e}(r,w),\sigma^2_{\scriptscriptstyle e}(r,w)\big), 
\end{equation}

\noindent then the theoretical model in (\ref{ch3:eq:sensingmodel}) can be rewritten as follows:
\begin{align} \label{ch3:eq:sensingmodel2} 
y_{it}(r,& w) = \; \mu_{\scriptscriptstyle K} \hspace{-0.5mm} \cdot \hspace{-0.5mm} \mu_{\scriptscriptstyle e}(r,w) + dK_i \hspace{-0.5mm} \cdot \hspace{-0.5mm} \mu_{\scriptscriptstyle e}(r,w) \nonumber \\
& + \mu_{\scriptscriptstyle K} \! \cdot \! de_{it}(r,w) + \mu_{\scriptscriptstyle K} \hspace{-0.5mm} \cdot \hspace{-0.5mm} D_{i} + \mu_{\scriptscriptstyle K} \hspace{-0.5mm} \cdot \hspace{-0.5mm} C_{t}(r,w) \nonumber \\
& + \mu_{\scriptscriptstyle K} \hspace{-0.5mm} \cdot \hspace{-0.5mm} R_{it} +  A_{it}, 
\end{align}

\noindent where $dK_i$ and $de_{it}(r,w)$ are the remaining zero-mean normal variables after removing their means, $\mu_{\scriptscriptstyle K}$ and $\mu_{\scriptscriptstyle e}(r,w)$, from the variables $K_i$ and $e_{it}(r,w)$ in equation (\ref{ch3:eq:sensingmodel}), respectively:
\begin{equation} \label{ch3:eq:dK}
  dK_{i} \sim  N (0,\sigma^2_{\scriptscriptstyle K}),
\end{equation}
\vspace{-6mm}
\begin{equation} \label{ch3:eq:de}
  de_{it}(r,w)  \sim N \big(0,\sigma^2_{\scriptscriptstyle e}(r,w)\big).
\end{equation}

\noindent In the previous equation (\ref{ch3:eq:sensingmodel2}), the component $dK_i \! \cdot \! de_{it}(r,w)$ has not been taken into consideration because it yields a negligible component.

Thus, the model in (\ref{ch3:eq:sensingmodel2}) will be the model to be approached by means of the proposed Bayesian multilevel random-effects model. The theoretical model in (\ref{ch3:eq:sensingmodel2}) consists in several and interacting random components dependent on reflectance and wavelength. The Bayesian approach by its properties of defining probability distributions for all the parameters and conditional dependencies among parameters in a fully probabilistic model, with fully propagation of uncertainty among parameters, is a suitable modeling framework for approaching the theoretical model in (\ref{ch3:eq:sensingmodel2}) with several and interacting random effects. The hierarchical (multilevel) structure of the proposed model arises from the conditional dependencies of the noise parameters on reflectance and wavelength.

Reflectance $r$ and wavelength $w$ are continuous factors in the theoretical model in (\ref{ch3:eq:sensingmodel2}). However, in practical experimentations wavelength is provided by the color band (wavelength range) of a color image and is usually treated as a categorical variable. Also, in our experimentation the real reflectance values of measured patches of the ColorChecker are unknown, so a convenient way to consider reflectance is as levels of a categorical variable.

Let us assume there is an array of observations $\bm{y} \in {\rm I\!R}^{N \hspace{-0.2mm} \times \hspace{-0.2mm} T \hspace{-0.2mm} \times \hspace{-0.2mm} R \hspace{-0.2mm} \times \hspace{-0.2mm} W}$ of image digital numbers, with an element $y_{it}(r,w)$ representing an observation of the image digital number registered at the $i$'th pixel and at the $t$'th image shoot and as a function of levels $r$ and $w$.
Similarly to the previous Section \ref{ch3:sensingmodel}, $N$ denotes the pixels in an image sensor ($i=1,...,N$), $T$ denotes the number of image exposures ($t=1,...,T$), $R$ denotes the number of levels of reflectance examined ($r=1,...,R$) and $W$ denotes the levels of wavelength examined ($w=1,...,W$).

The collection $\bm{y}$ of observations is considered to follow a Normal distribution depending on an underlying mean function $\bm{f}$ and standard deviation of the noise $\sigma$,
\begin{equation} \label{ch3:eq:model}
p(\bm{y}|\bm{f}) = \mathcal{N}(\bm{y}|\bm{f},\sigma^2 \bm{I}),
\end{equation}

\noindent where $\bm{I}$ is the identity matrix. The mean function $\bm{f}$ is a sum function of independent random effects nested inside the fixed effects of the categorical variables $r$ and $w$. Thus, for a single observation $(i,t)$, the underlying function takes the form: 
\begin{align} \label{ch3:eq:underlying_model}
  f_{it}(r,w) =& \; \mu_{\scriptscriptstyle 0}(r,w) + S_i(r,w) + F_i + T_t(r,w)
  \nonumber \\
  &+ P_{it}(r,w). 
\end{align}

\begin{figure}
\centering
  \includegraphics[trim = 27mm 0mm 42mm 0mm, clip, width=\columnwidth]{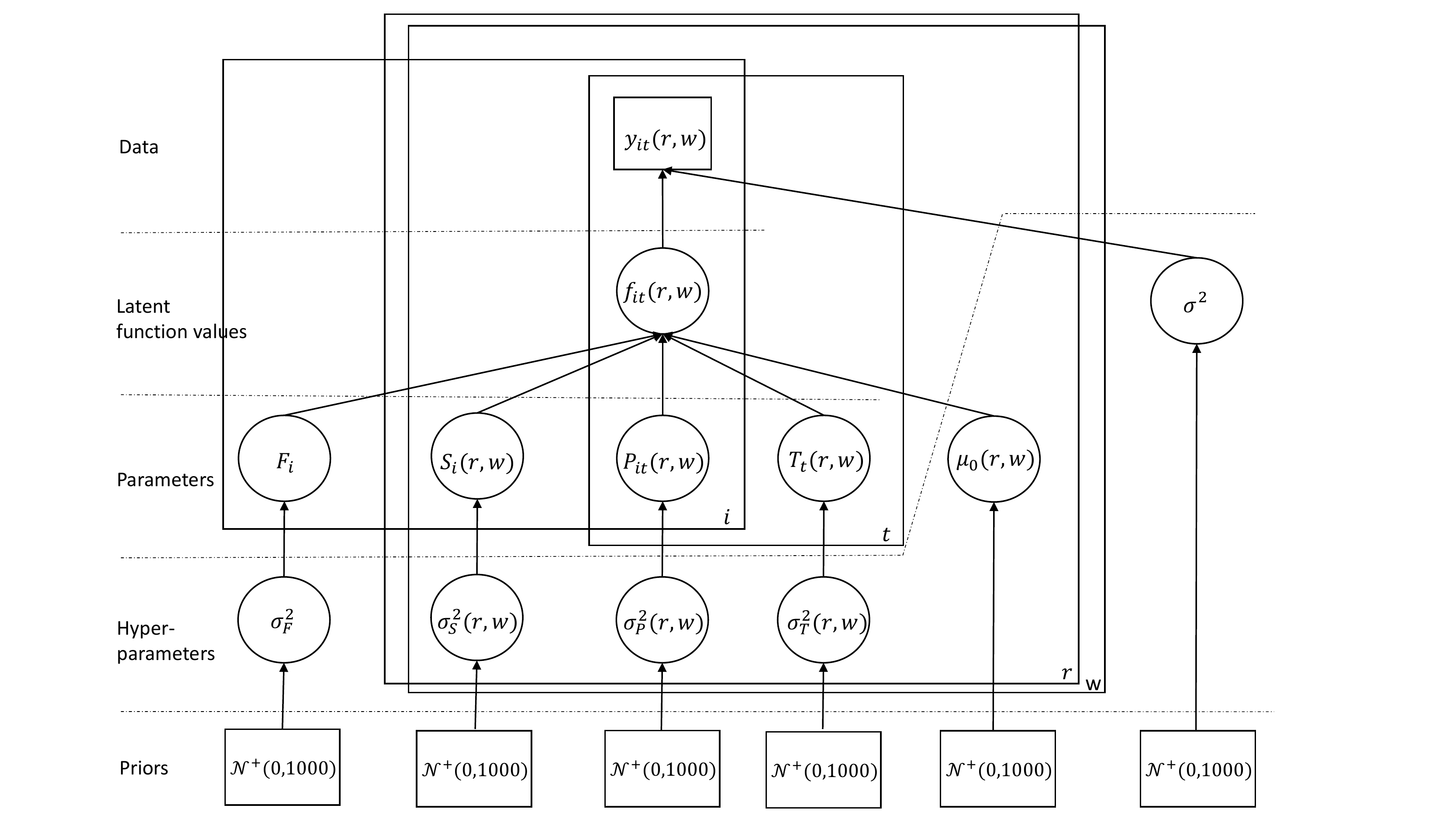}
\caption{Directed acyclic graph of the proposed Bayesion model in equations (\ref{ch3:eq:model}) and (\ref{ch3:eq:underlying_model}).}
\label{ch3_directed_acyclic_graph}  
\end{figure}

\noindent In Figure \ref{ch3_directed_acyclic_graph} the directed acyclic graph of the proposed Bayesian model in equations (\ref{ch3:eq:model}) and (\ref{ch3:eq:underlying_model}) is depicted. Parameter $\mu_{\scriptscriptstyle 0}(r,w)$ is the fixed effect of the specific levels $r$ and $w$ of reflectance and wavelength, respectively. It gathers component $\mu_{\scriptscriptstyle K} \cdot \mu_{\scriptscriptstyle e}(r,w)$ in the theoretical model (\ref{ch3:eq:sensingmodel2}), which represents the mean reflectance as a function of $r$ and $w$.

Parameter $S_i(r,w)$ in (\ref{ch3:eq:underlying_model}) models component $dK_i \! \cdot \! \mu_{\scriptscriptstyle e}(r,w)$ in the theoretical model (\ref{ch3:eq:sensingmodel2}), where $\mu_{\scriptscriptstyle e}(r,w)$ is the mean electrons as a function of reflectance $r$ and wavelength $w$, and $dK_i$ is the PRNU zero-mean normal random variable (\ref{ch3:eq:dK}). Therefore, $S_{i}(r,w)$ can be modeled as a zero-mean normal prior distribution, defined on the pixel dimension $i$ and as a function of $r$ and $w$, 
\begin{equation} \label{ch3:eq:S}
p \bigl( S_{i}(r,w)|\sigma_{\scriptscriptstyle S}(r,w) \bigr) = \mathcal{N} \bigl( S_{i}(r,w)|0,\sigma^2_{\scriptscriptstyle S}(r,w) \bigr).
\end{equation}

\noindent The standard deviation $\sigma_{\scriptscriptstyle S}(r,w)$ of this parameter $S$ models the PRNU for specific levels $r$ and $w$.

Parameter $F_i$ in (\ref{ch3:eq:underlying_model}) models component $\mu_{\scriptscriptstyle K} \! \cdot \! D_i$ in the theoretical model (\ref{ch3:eq:sensingmodel2}), where $\mu_{\scriptscriptstyle K}$ is a constant and $D_i$ is Poisson distributed (\ref{ch3:eq:FPN}). The assumption of considering Poisson generated electrons after their conversion to digital numbers ($\mu_{\scriptscriptstyle K} \! \cdot \! D_i$) to normally distributed variables is truly reasonable in this context. Therefore, the parameter $F_i$ in (\ref{ch3:eq:underlying_model}) is modeled following a zero-mean Normal prior distribution:
\begin{equation} \label{ch3:eq:F}
p \bigl( F_i|\sigma_{\scriptscriptstyle F} \bigr) = \mathcal{N} \bigl( F_i|0,\sigma_{\scriptscriptstyle F}^2 \bigr),
\end{equation}

\noindent whose standard deviation $\sigma_{\scriptscriptstyle F}$ represents the FPN (\ref{ch3:eq:FPN}) of the image sensor, which does not depend on reflectance or wavelength.

The parameter $T_t(r,w)$ in (\ref{ch3:eq:underlying_model}) models component $\mu_{\scriptscriptstyle K} \! \cdot \! C_t(r,w)$ in the theoretical model (\ref{ch3:eq:sensingmodel2}), where $\mu_{\scriptscriptstyle K}$ is a constant and $C_t(r,w)$ is the current noise Poisson variable (\ref{ch3:eq:Current}). Like in the previous case of $\mu_{\scriptscriptstyle K} \! \cdot \! D_i$, the Poisson variable $\mu_{\scriptscriptstyle K} \! \cdot \! C_t(r,w)$ can be approximated as a Normal prior distribution by the parameter $T_t(r,w)$ as see below:
\begin{equation} \label{ch3:eq:T}
p \bigl( T_t(r,w)|\sigma_{\scriptscriptstyle T}(r,w) \bigr) = \mathcal{N} \bigl( T_t(r,w)|0,\sigma^2_{\scriptscriptstyle T}(r,w) \bigr).
\end{equation}

\noindent The standard deviation $\sigma_{\scriptscriptstyle T}(r,w)$ of this parameter $T$ will represent the current noise for specific levels $r$ and $w$.

The parameter $P_{it}(r,w)$ in (\ref{ch3:eq:underlying_model}) models component $\mu_{\scriptscriptstyle K} \! \cdot \! de_{it}(r,w)$ in the theoretical model (\ref{ch3:eq:sensingmodel2}), where $\mu_{\scriptscriptstyle K}$ is a constant and $de_{it}(r,w)$ is a zero-mean Normal variable (\ref{ch3:eq:de}). Then, $P_{it}(r,w)$ is modeled as a zero-mean Normal variable:
\begin{equation} \label{ch3:eq:P}
p \bigl( P_{it}(r,w)|\sigma_{\scriptscriptstyle P}(r,w) \bigr) = \mathcal{N} \bigl( P_{it}(r,w)|0,\sigma^2_{\scriptscriptstyle P}(r,w) \bigr),
\end{equation}

\noindent where its variance $\sigma^2_{\scriptscriptstyle P}(r,w)$ represents the photon noise for specific levels $r$ and $w$. 

Finally, the residual of the model in (\ref{ch3:eq:model}) will gather components $\mu_{\scriptscriptstyle K} \cdot R_{it}$ and $A_{it}$ in the theoretical model (\ref{ch3:eq:sensingmodel2}) which are expected to be random Normal variables defined independently on both dimensions $i$ and $t$ in equations (\ref{ch3:eq:Reset}) and (\ref{ch3:eq:Ampli}). These residuals will also contain other possible independent and random uncontrolled noise factors in the experimentation or even in the process.
\vspace{0.3cm}

The likelihood function of the observations $\bm{y}$ given the parameters $\bm{\mu}_{\scriptscriptstyle 0}=\{\mu_{\scriptscriptstyle 0}(r,w)\}$, $\bm{S}=\{S_i(r,w)\}$, $\bm{F}=\{F_i\}$, $\bm{T}=\{T_t(r,w)\}$, $\bm{P}=\{P_{it}(r,w)\}$, and $\sigma$, is written in equation (\ref{ch3:eq:likeli}).
\begin{figure*}[!t]
\begin{align} \label{ch3:eq:likeli}
p ( \bm{y}|\bm{\mu}_{\scriptscriptstyle 0},\bm{S},\bm{F},\bm{T},\bm{P},\sigma ) = 
\prod_{\forall i,t,r,w} \mathcal{N} \bigl( y_{it}(r,w)|\mu_{\scriptscriptstyle 0}(r,w),S_i(r,w),F_i,T_t(r,w),P_{it}(r,w),\sigma \bigr).
\end{align}
\end{figure*}

\subsection{Bayesian inference} \label{ch3:bayesinference} 

Bayesian inference is done over the joint posterior distribution of parameters and hyperparameters given the data, which is proportional to the likelihood and priors. 

The joint posterior distribution of the proposed model is written in equation (\ref{ch3:eq:posterior}),
\begin{figure*}[!t]
\begin{align} \label{ch3:eq:posterior}
p(\bm{\mu_{\scriptscriptstyle 0}}&,\bm{S},\bm{F},\bm{T},\bm{P},\sigma|\bm{y}) \propto   \nonumber \\
 &  p(\bm{y}|\bm{\mu_{\scriptscriptstyle 0}},\bm{S},\bm{F},\bm{T},\bm{P},\sigma) \, p(\bm{\mu_{\scriptscriptstyle 0}}) \, p(\bm{S}|\bm{\sigma}_{\scriptscriptstyle S}) \, p(\bm{F}|\sigma_{\scriptscriptstyle F}) \, p(\bm{T}|\bm{\sigma}_{\scriptscriptstyle T}) \, p(\bm{P}|\bm{\sigma}_{\scriptscriptstyle P}) \, p(\sigma)p(\bm{\sigma}_{\scriptscriptstyle S})
 p(\sigma_{\scriptscriptstyle F}) \, p(\bm{\sigma}_{\scriptscriptstyle T}) \, p(\bm{\sigma}_{\scriptscriptstyle P}) \propto \nonumber \\ 
& \prod_{\forall i,t,r,w} \mathcal{N}\bigl( y_{it}(r,w)|\mu_{\scriptscriptstyle 0}(r,w),S_i(r,w),F_i,T_t(r,w) ,P_{it}(r,w),\sigma \bigr)   
\times \mathcal{N}\bigl(\mu_{\scriptscriptstyle 0}(r,w)|0,1000\bigr)     \nonumber \\  
& \times \prod_{\forall i,r,w} \mathcal{N} \bigl( S_i(r,w)|0,\sigma^2_{\scriptscriptstyle S}(r,w) \bigr) \prod_{\forall i,r,w}  \mathcal{N}\bigl(F_i|0,\sigma_{\scriptscriptstyle F}^2 \bigr) \prod_{\forall i,r,w} \mathcal{N} \bigl( T_t(r,w)|0,\sigma^2_{\scriptscriptstyle T}(r,w) \bigr)  \nonumber \\
&\times  \prod_{\forall i,r,w} \mathcal{N}\bigl(P_{it}(r,w)|0,\sigma^2_{\scriptscriptstyle P}(r,w) \bigr) \, \mathcal{N} \bigl( \sigma|0,1000 \bigr) \, \mathcal{N} \bigl( \sigma_{\scriptscriptstyle S}(r,w)|0,1000 \bigr) \, \mathcal{N} \bigl( \sigma_{F}|0,1000 \bigr)    \nonumber \\
& \times \mathcal{N} \bigl( \sigma_{\scriptscriptstyle T}(r,w)|0,1000 \bigr) \, \mathcal{N} \bigl( \sigma_{\scriptscriptstyle P}(r,w)|0,1000 \bigr)
\end{align}
\end{figure*}
where $p(\bm{y}|\bm{\mu_{\scriptscriptstyle 0}},\bm{S},\bm{F},\bm{T},\bm{P},\sigma)$ is the likelihood of the model in (\ref{ch3:eq:likeli}), and $p(\bm{S}|\bm{\sigma}_{\scriptscriptstyle S})$, $p(\bm{F}|\sigma_{\scriptscriptstyle F})$, $p(\bm{T}|\bm{\sigma}_{\scriptscriptstyle T})$, and $p(\bm{P}|\bm{\sigma}_{\scriptscriptstyle P})$ the priors for the corresponding parameters in (\ref{ch3:eq:S}), (\ref{ch3:eq:F}), (\ref{ch3:eq:T}) and (\ref{ch3:eq:P}), respectively, and $p(\bm{\mu}_{\scriptscriptstyle 0})$, $p(\sigma)$, $p(\bm{\sigma_{\scriptscriptstyle S})}$, $p(\sigma_{\scriptscriptstyle F})$, $p(\bm{\sigma_{\scriptscriptstyle T})}$ and $p(\bm{\sigma_{\scriptscriptstyle P})}$ the priors for the hyperparameters, where $\bm{\sigma}_{\scriptscriptstyle S}$ denotes the collection $\{\sigma_{\scriptscriptstyle S}(r,s)\}$, and similarly $\bm{\sigma}_{\scriptscriptstyle T}=\{\sigma_{\scriptscriptstyle T}(r,s)\}$ and $\bm{\sigma}_{\scriptscriptstyle P}=\{\sigma_{\scriptscriptstyle P}(r,s)\}$.
If no prior information is available for the hyperparameters, vague prior distributions still need to be specified. For parameters $\bm{\mu}_{\scriptscriptstyle 0}$, vague Normal distributions with large variances are defined. For the standard deviation parameters $\bm{\sigma}_{\scriptscriptstyle S}$, $\sigma_{\scriptscriptstyle F}$, $\bm{\sigma}_{\scriptscriptstyle T}$, $\bm{\sigma}_{\scriptscriptstyle P}$ and $\sigma$, positive half-Normal distributions with large variances \citep{kass1995reference,yang1996catalog} are used.

The joint posterior distribution of the parameters have been estimated with MCMC using Gibbs sampling \citep{geman1993stochastic,brooks_2011} and the WinBUGS software \citep{lunn2000winbugs,ntzoufras2011bayesian}. Samples from the joint posterior distribution of the model parameters are obtained, and estimates and credible intervals are inferred for the model parameters. Three simulation chains have been launched for every one of the parameters, with 100000 iterations, of which the first 30000 iterations were rejected as burn-in, and finally, only 1 of every 100 was retained with the aim of reducing the correlation in the samples. The convergence of the simulation chains was evaluated with the split-Rhat convergence diagnosis and the effective sample size of the chains \citep{gelman1992inference,vehtari2019ranknormalization}. A value of 1 in the split-Rhat convergence statistic indicates good mixing of simulated chains. Accepted good values for the split-Rhat statistic would be between 1 and 1.1, although a more strict range has also been suggested recently \citep{vehtari2019ranknormalization}. In this study, values of the split-Rhat statistic lower than 1.05 have been obtained for all parameters.

\section{Experimental results and analysis}\label{ch3:results}

In this work, the interest is in analyzing the standard deviation parameters $\bm{\sigma}_{\scriptscriptstyle S}$, $\sigma_{\scriptscriptstyle F}$, $\bm{\sigma}_{\scriptscriptstyle T}$, $\bm{\sigma}_{\scriptscriptstyle P}$ and $\sigma$, which are the quantities that allow us to characterize the mean noise caused by the parameters $\bm{S}$, $\bm{F}$, $\bm{T}$, $\bm{P}$ and $residuals$, respectively. They are in units of output digital numbers.

As it will be seen below, some of the noise estimates are reflectance dependent, fact that suggests the computation of their coefficients of variation, in which the linear effect of reflectance (linear-multiplicative effect of the mean number of electrons) on the parameters is removed. In this way, different sensors or different experimentations with different dynamic ranges can be compared.

The coefficient of variation ($CV$) is the ratio between the standard deviation and the mean of the component considered ($CV=\sigma / \mu$), that is, the inverse of the signal-to-noise ratio. In fact, the coefficient of variation defines the quality of a sensor as a discriminatory power of a signal. The overall means are represented by $\bm{\mu_{\scriptscriptstyle 0}}$.

\subsection{Standard deviation of the parameters} \label{ch3:std}

\begin{figure*}[!t]
\centering
\begin{tabular}{ccc}
{ (a) \hspace{0.4mm} $\bm{\sigma}_{\scriptscriptstyle S}$} & { (b) \hspace{0.4mm} $\sigma_{\scriptscriptstyle F}$} & { (c) \hspace{0.4mm} $\bm{\sigma}_{\scriptscriptstyle T}$} \\
\includegraphics[trim = 10mm 0mm 10mm 20mm, clip, scale=0.34]{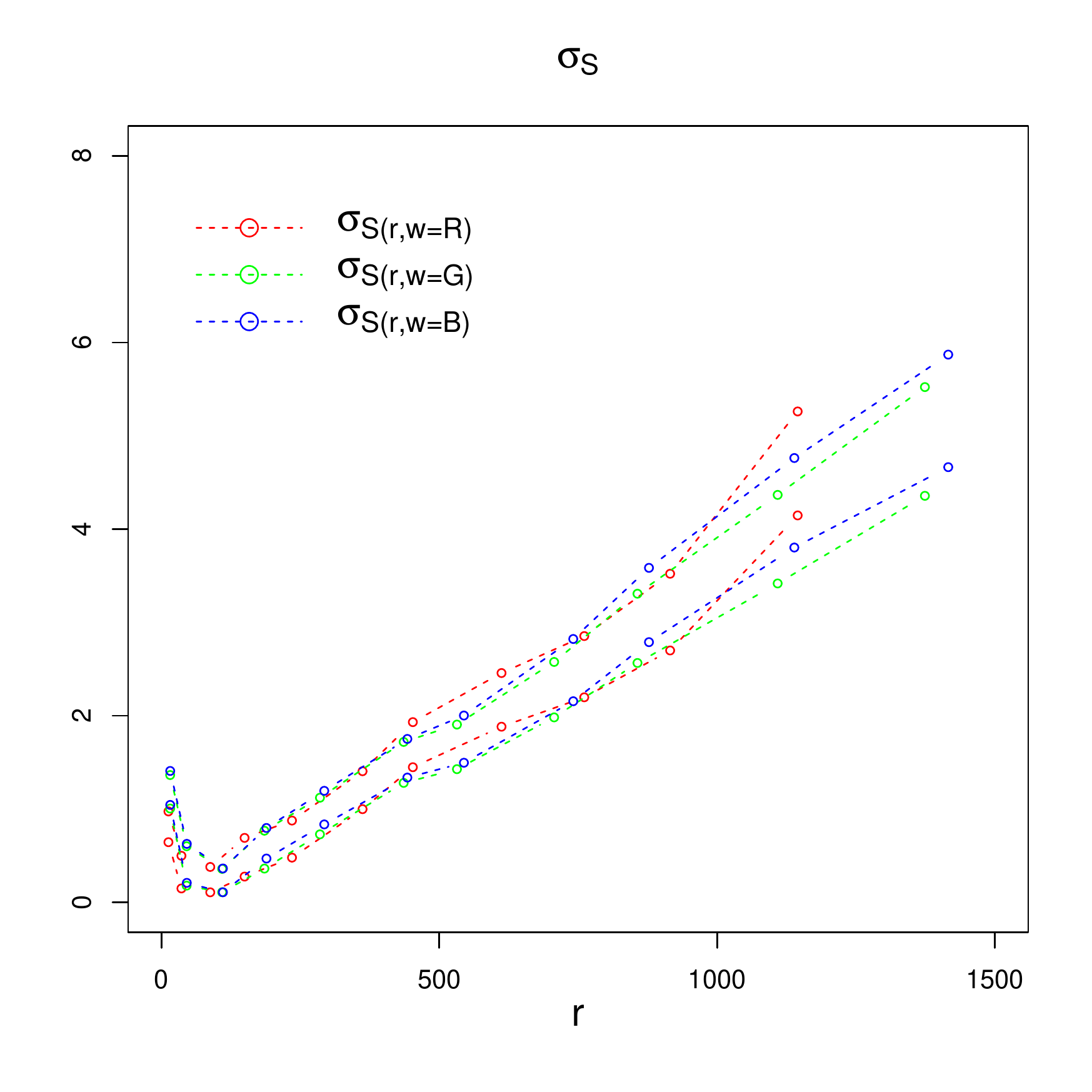}
& \includegraphics[trim = 10mm 0mm 10mm 20mm, clip, scale=0.34]{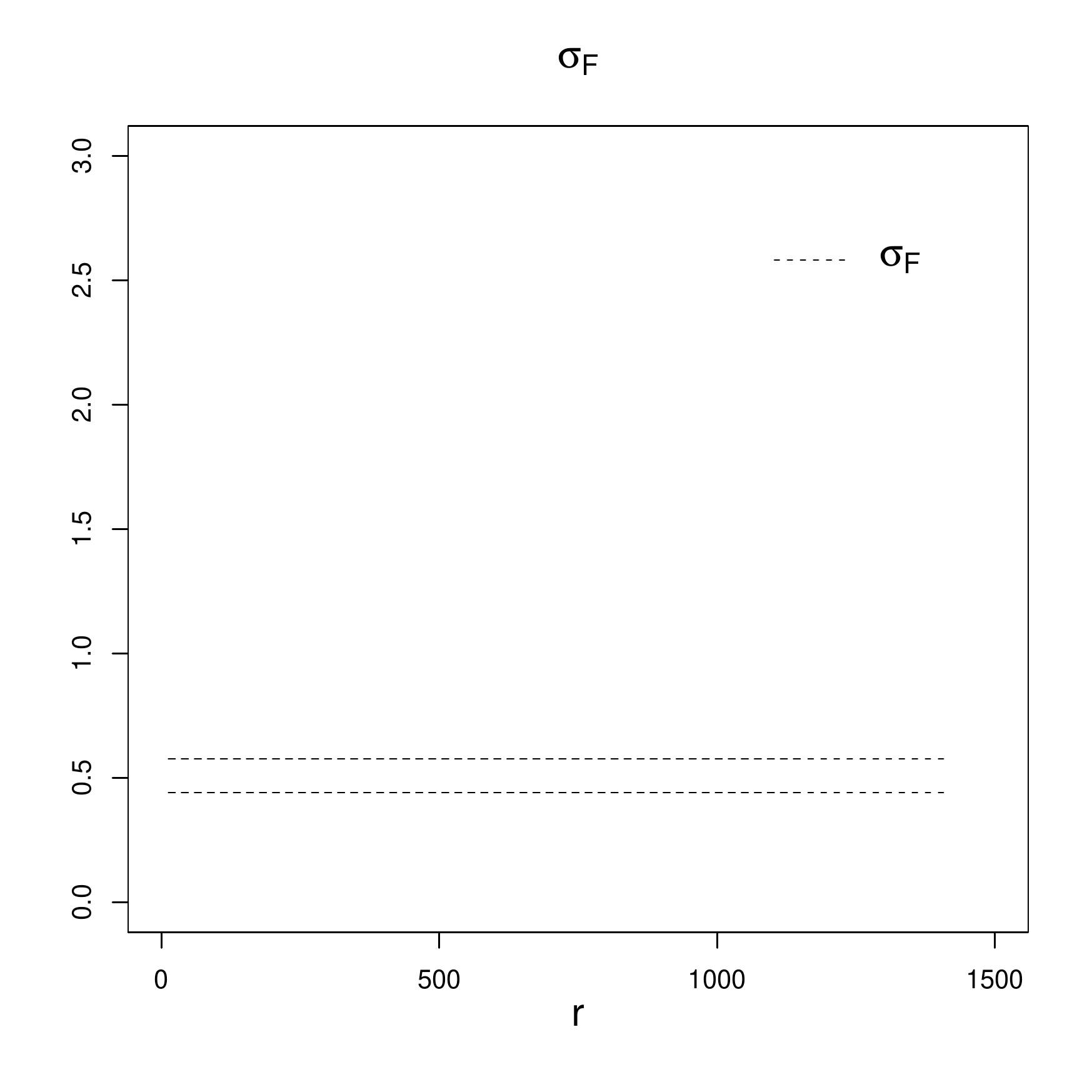}
& \includegraphics[trim = 10mm 0mm 10mm 20mm, clip, scale=0.34]{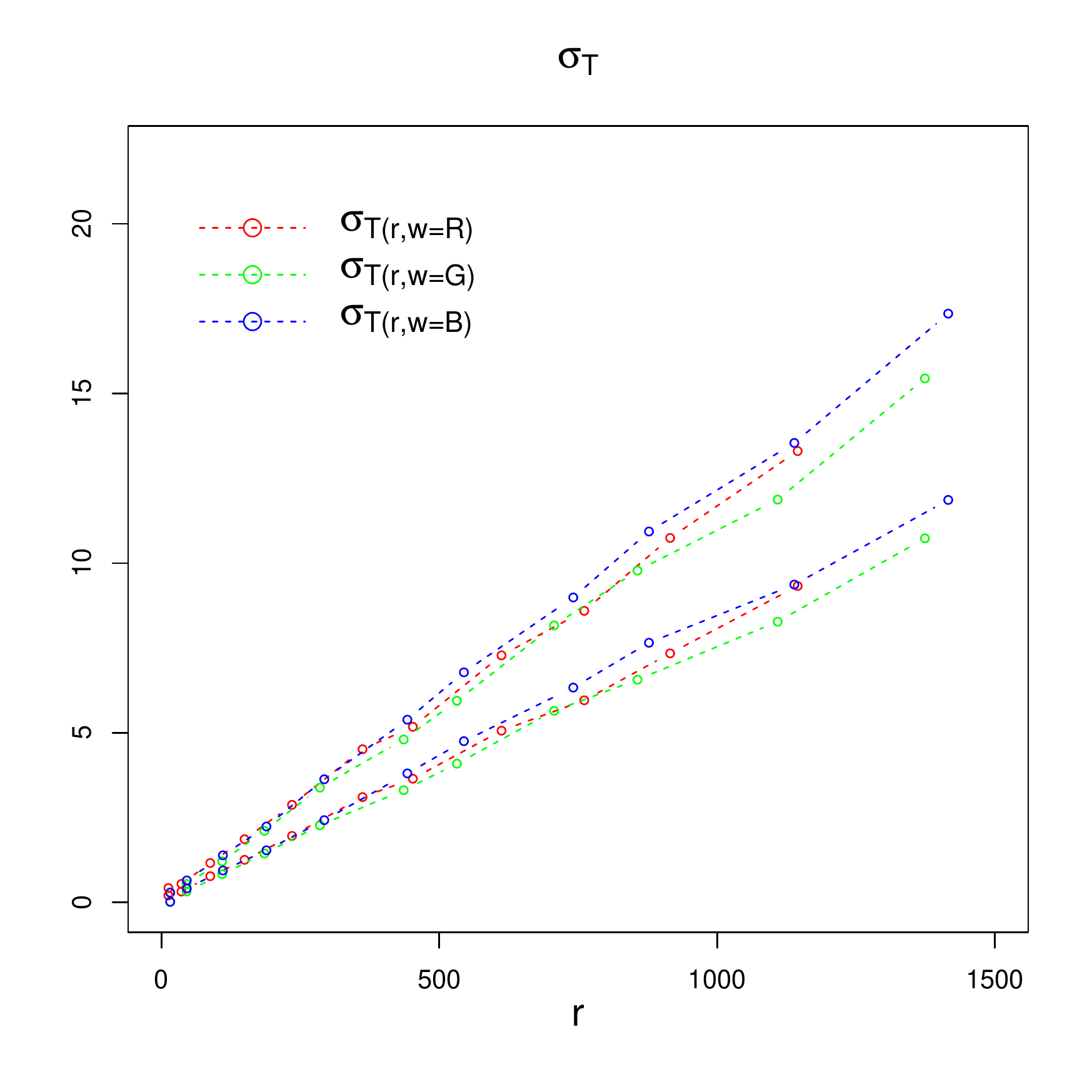}
\vspace{0.03cm} \\
\multicolumn{3}{c} {\begin{tabular}{cc} {(d) \hspace{0.4mm} $\bm{\sigma}_{\scriptscriptstyle P}$} & {(e) \hspace{0.4mm} $\bm{\sigma}$}
\\
\includegraphics[trim = 10mm 0mm 0mm 20mm, clip, scale=0.34]{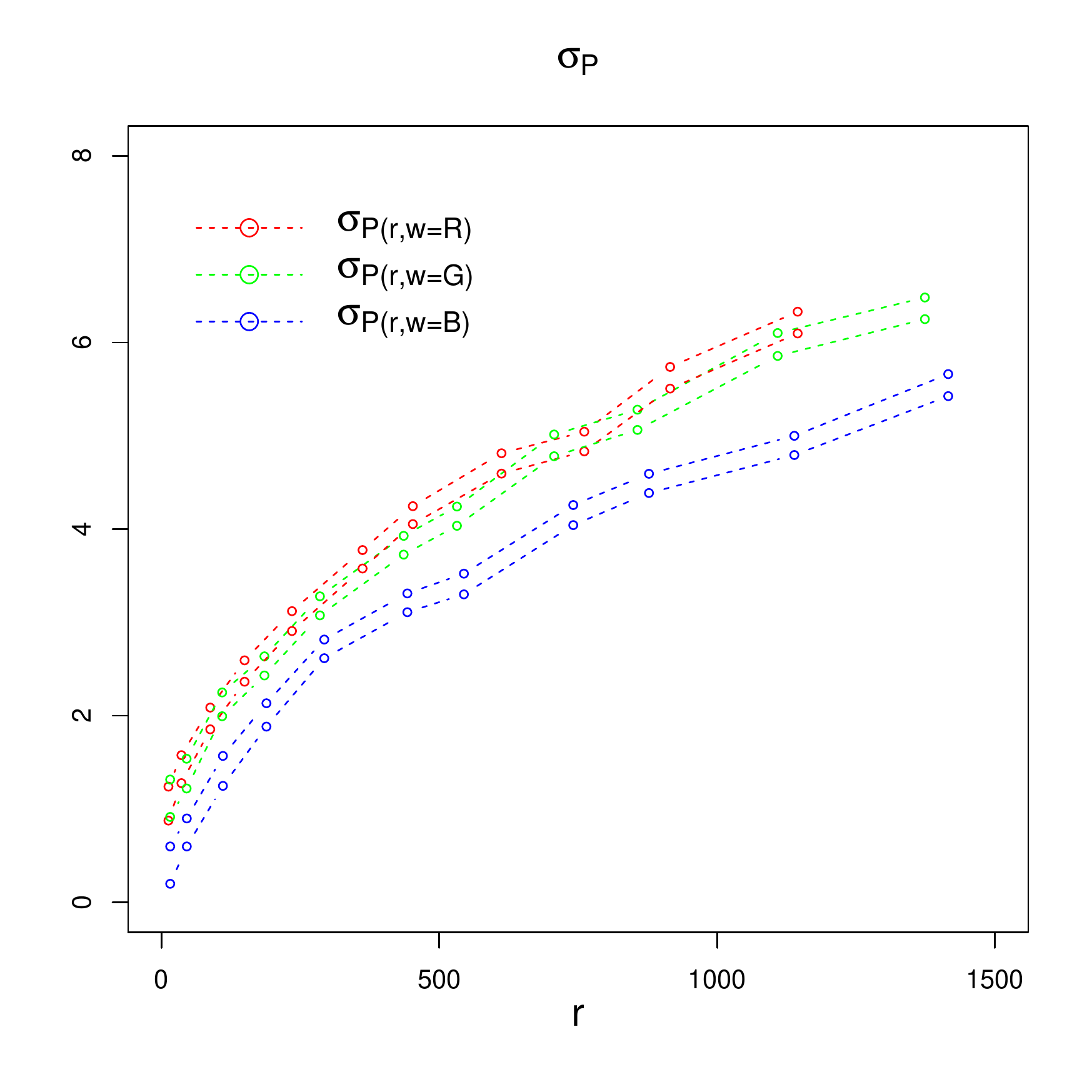} &
\includegraphics[trim = 10mm 0mm 0mm 20mm, clip, scale=0.34]{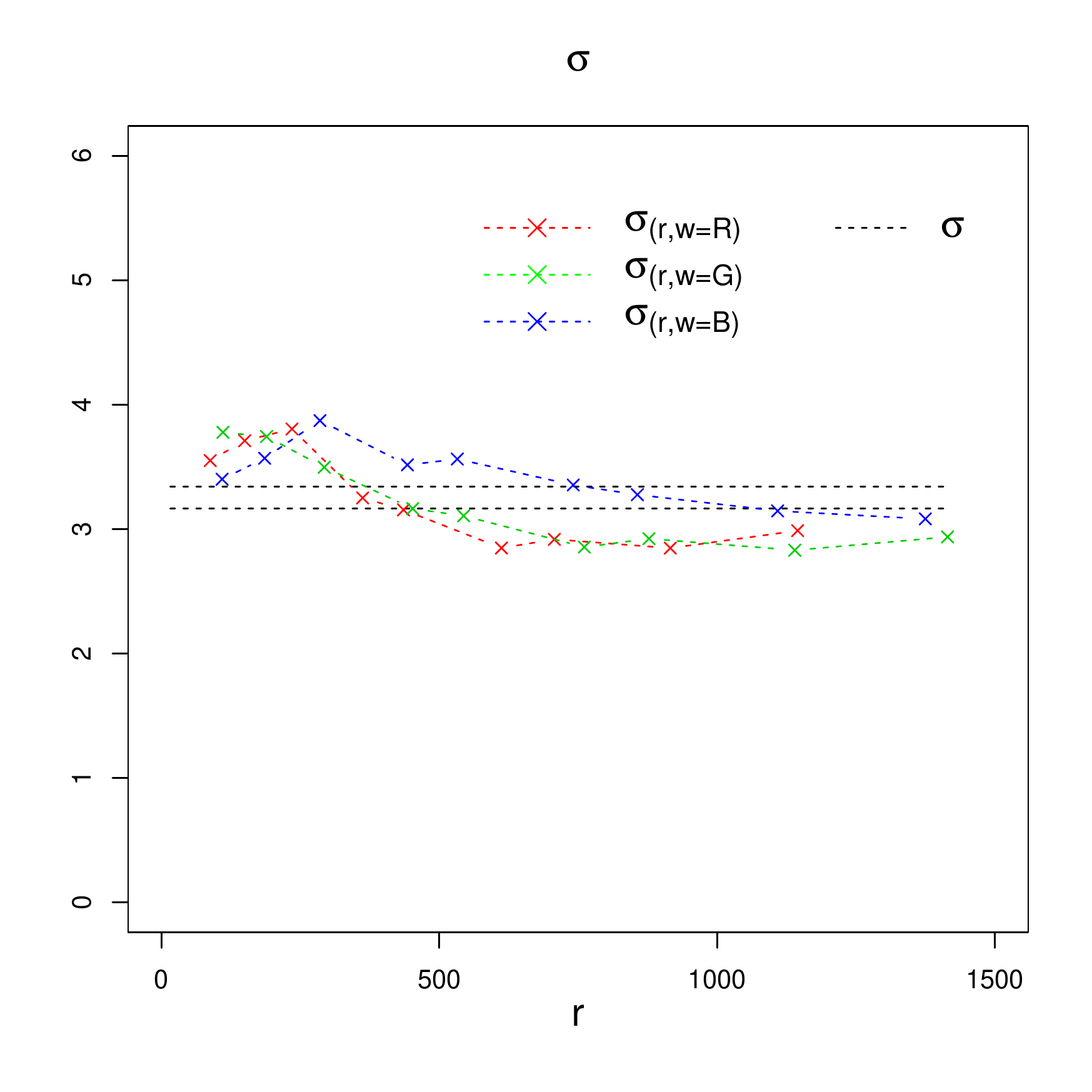}
\end{tabular}  }
\end{tabular}
\vspace{-4mm}
\caption{2.5\% and 97.5\% credible intervals for the standard deviation of the parameters $S$ ($\bm{\sigma}_{\scriptscriptstyle S}$) (a), $F$ ($\sigma_{\scriptscriptstyle F}$) (b), $T$ ($\bm{\sigma}_{\scriptscriptstyle T}$) (c), $P$ ($\bm{\sigma}_{\scriptscriptstyle P}$) (d), and  $residuals$ ($\sigma$) (e), versus mean output-reflectance $r$ and wavelengths $w$. In (e), the residual deviation as a function of reflectance $r$ and wavelength $w$ ($\sigma_{(r,w)}$) is computed and plotted jointly with the mean residual deviation ($\sigma$).}
\label{ch3_fig03}
\end{figure*}

Figure \ref{ch3_fig03} shows the 95\% pointwise credible intervals for parameters $\bm{\sigma}_{\scriptscriptstyle S}$ (PRNU), $\sigma_{\scriptscriptstyle F}$ (FPN), $\bm{\sigma}_{\scriptscriptstyle T}$ (current noise), $\bm{\sigma}_{\scriptscriptstyle P}$ (photon noise) and $\sigma$ (reset, amplifier, flicker and quantization noises). They are plotted against the mean effects of the reflectance and wavelength variables which are modeled by parameters $\mu_{\scriptscriptstyle 0}(r,w)$.

As stated in Section \ref{ch3:proposed}, parameter $\sigma_{\scriptscriptstyle S}(r,w)$ models the noise effects of component $\mu_{\scriptscriptstyle e}(r,w) \cdot dK_i$ in the theoretical model (\ref{ch3:eq:sensingmodel2}), and represents the mean noise caused by the interpixel differences when generating electrons from the incoming photons, effect called PRNU and encapsulated in the gain random variable $dK_i$; see equations (\ref{ch3:eq:PRNU}) and (\ref{ch3:eq:dK}). The increasing of $\bm{\sigma}_{\scriptscriptstyle S}$ with respect to reflectance $r$ ($x$-axis) that can be seen in Figure \ref{ch3_fig03}(a) is due to the linear-multiplicative effect of the electrons $\mu_{\scriptscriptstyle e}(r,w)$ on $dK_i$, since $dK_i$ is expected to be a zero-mean normal variable independent on reflectance; see equation (\ref{ch3:eq:dK}). However, this linear behaviour that can be appreciated in the figure is broken at the lowest values of reflectance. In fact, a non-linear interaction between PRNU and the light intensity in low and high illumination levels has been pointed out \citep{gow2007comprehensive}. Estimates for this effect range from around 0.2 digital numbers at low reflectances, to around 6 digital numbers at the highest reflectances examined in the experimentation.

The estimated parameter $\sigma_{\scriptscriptstyle F}$ is not dependent on reflectance and was estimated around 0.5 digital numbers; see Figure \ref{ch3_fig03}(b). This represents the mean noise, whose standard deviation is in equation (\ref{ch3:eq:F}), caused by Poisson distributed dark current variations across pixels (FPN; see equation (\ref{ch3:eq:FPN})), an effect without the need of incident light. The estimate found for this effect can be considered a negligible value for the Foveon X3\textregistered~ image sensor, as specified in the characteristics provided by the manufacturer.

The estimated parameter $\sigma_{\scriptscriptstyle T}(r,w)$ shows a linear dependency with respect to reflectance $r$, either in mean and in variance; see Figure \ref{ch3_fig03}(c). This linear behaviour was expected since it represents the mean noise, whose standard deviation appears in equation (\ref{ch3:eq:T}), caused by Poisson distributed free electrons thermally generated during the exposure time (current noise; see equation (\ref{ch3:eq:Current})), and temperature inside a pixel, at a given exposure time, is directly related to incident light and therefore to reflectance as well. Estimates for this effect range from approximately 0 digital numbers at the lowest reflectances, to around 15 digital numbers at the highest reflectances examined in the experimentation.

The estimated parameter $\sigma_{\scriptscriptstyle P}(r,w)$ does not increase linearly with respect to reflectance $r$ (Figure \ref{ch3_fig03}(d)). As stated in Section \ref{ch3:proposed}, $\sigma_{\scriptscriptstyle P}(r,w)$ models the noise effects of component $\mu_{\scriptscriptstyle K} \cdot de_{it}(r,w)$ in the theoretical model (\ref{ch3:eq:sensingmodel2}). $\mu_{\scriptscriptstyle K}$ is a constant and the electrons $de_{it}(r,w)$  are normal approximations (see equations (\ref{ch3:eq:photon2}) and (\ref{ch3:eq:de})) to Poisson variables (see equation (\ref{ch3:eq:electrons})), so that their standard deviation increases with the square root of the mean electrons $\bigl( \sqrt{\mu_{\scriptscriptstyle e}(r,w)} \bigr)$. Then, the slope of $\bm{\sigma}_{\scriptscriptstyle P}$ will be due to the variance of $de_{it}(r,w)$ (photon noise) which increases with the square root of electrons or, equivalently, with the squared root of reflectance. Thus, estimates for this effect increase proportionally to the square root of the reflectance from very low digital numbers at the lowest reflectances, to around 6 digital numbers at the highest reflectances, with slight differences among wavelengths.

Finally, the specifications of the sensor also indicate low-readout noise effects, for which and jointly with the reset noise, a mean error of 3.3 digital numbers was estimated in this study by the residual standard deviation $\sigma$ (Figure \ref{ch3_fig03}(e)). The residuals are not completely independent with respect to reflectance $r$, and a slight decreasing trend in residual deviation at low reflectance can be found. However, this lack of independence on the residuals is clearly very small with trend effects lower than 1 digital number and, hence, can be considered negligible in practice. For reflectance \hspace{-1.5mm} > \! 500, the residuals are without trend. This fact reflects that some of the noise components (reset, amplifier, flicker, and quantization noises) included in the residual deviation parameter might be slightly dependent on reflectance at low intensities.

Gain factor $\mu_{\scriptscriptstyle K}$ is embedded in all the noise parameter estimates, so they represent units of output digital numbers (electrons times gain factor). The differences among wavelengths reflect different behaviours, that is, the wavelengths R, G, and B do not generate exactly the same noise under the same conditions.

\subsection{Variation coefficients of the parameters} \label{ch3:CV}

Due to the dependency of the noise estimates $\bm{\sigma}_{\scriptscriptstyle P}$, $\bm{\sigma}_{\scriptscriptstyle T}$ and $\bm{\sigma}_{\scriptscriptstyle S}$ on the level of reflectance imaged (Figures \ref{ch3_fig03}(a), \ref{ch3_fig03}(c), and \ref{ch3_fig03}(d)), their coefficients of variation are computed. For parameters $\bm{F}$ and $residuals$ the computation of their variation coefficients makes no sense since parameter $\bm{F}$ is an independent variable on reflectance (\ref{ch3:eq:FPN}) and residuals can be considered in practice variance-constant with respect to reflectance (cf. Secction \ref{ch3:std}). Their absolute mean noise effects, $\sigma_{\scriptscriptstyle F}$ and $\sigma$, were estimated around 0.5 and 3.3 digital numbers, respectively; see Figures \ref{ch3_fig03}(b) and \ref{ch3_fig03}(e).

\begin{figure*}
\centering
\begin{tabular}{ccc}
{(a) \hspace{0.4mm} \small $CV_S$} & {(b) \hspace{0.4mm} \small $CV_T$} & {(c) \hspace{0.4mm} \small $CV_P$} \\
\includegraphics[trim = 10mm 0mm 10mm 20mm, clip, scale=0.34]{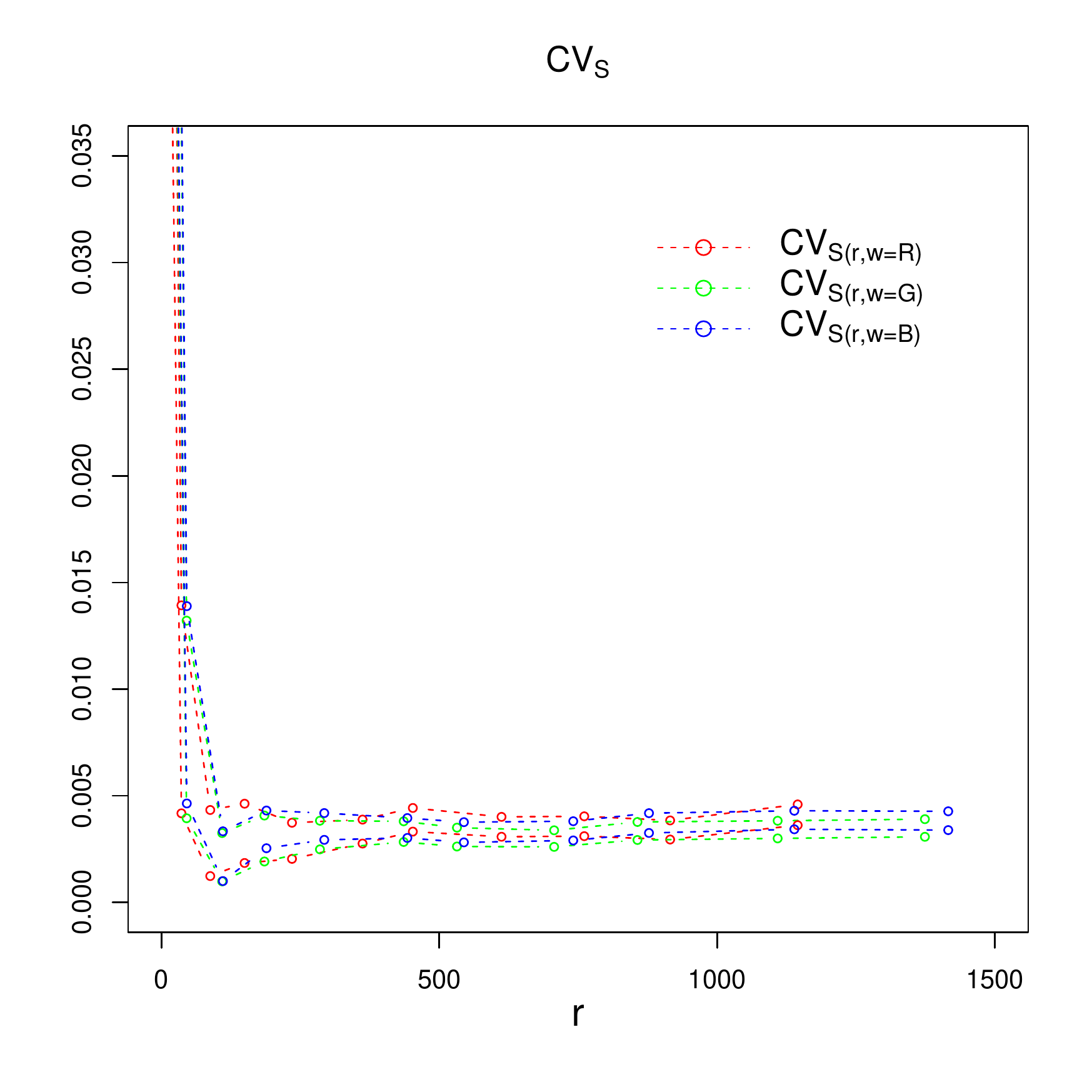} 
&\includegraphics[trim = 10mm 0mm 10mm 20mm, clip, scale=0.34]{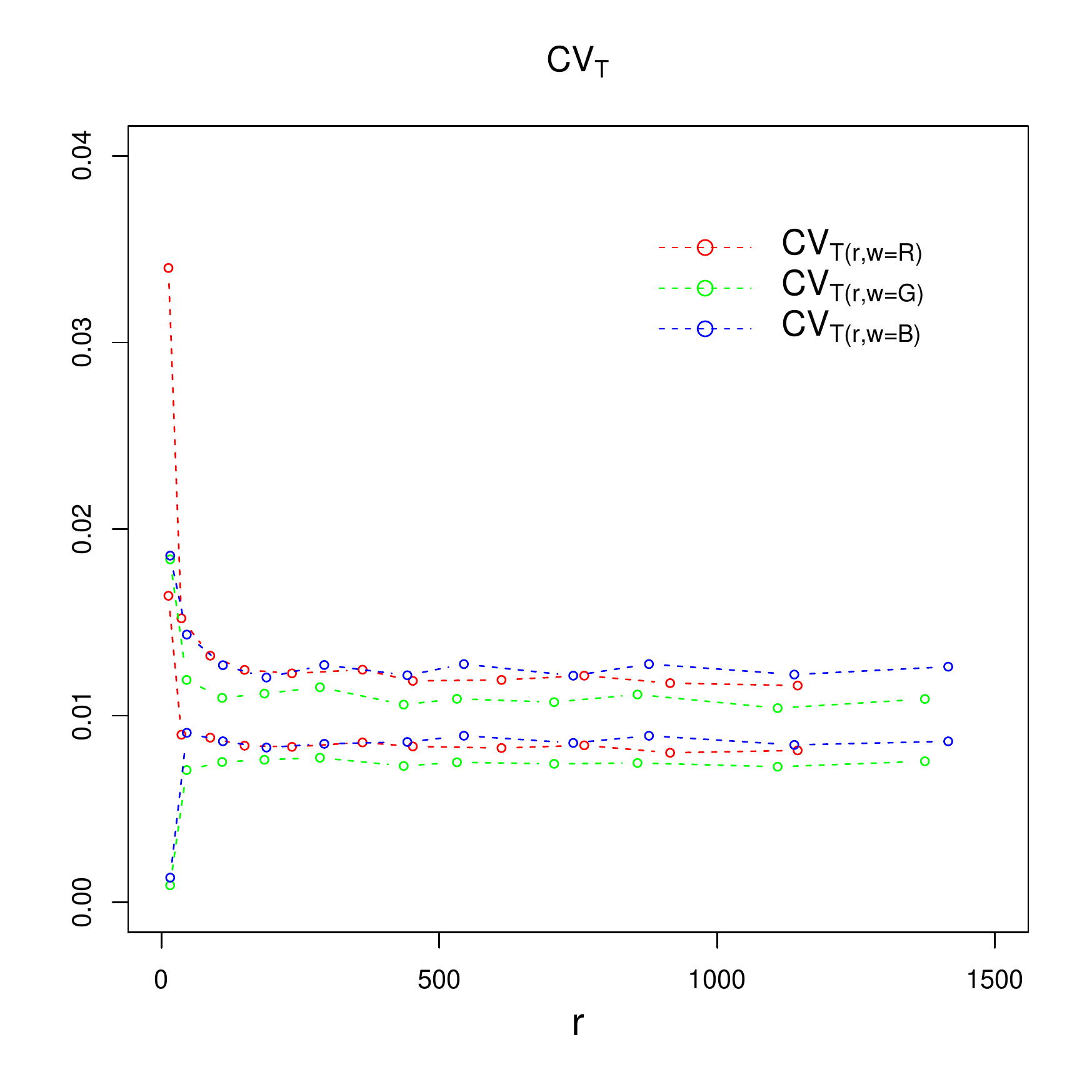}
 & \includegraphics[trim = 10mm 0mm 10mm 20mm, clip, scale=0.34]{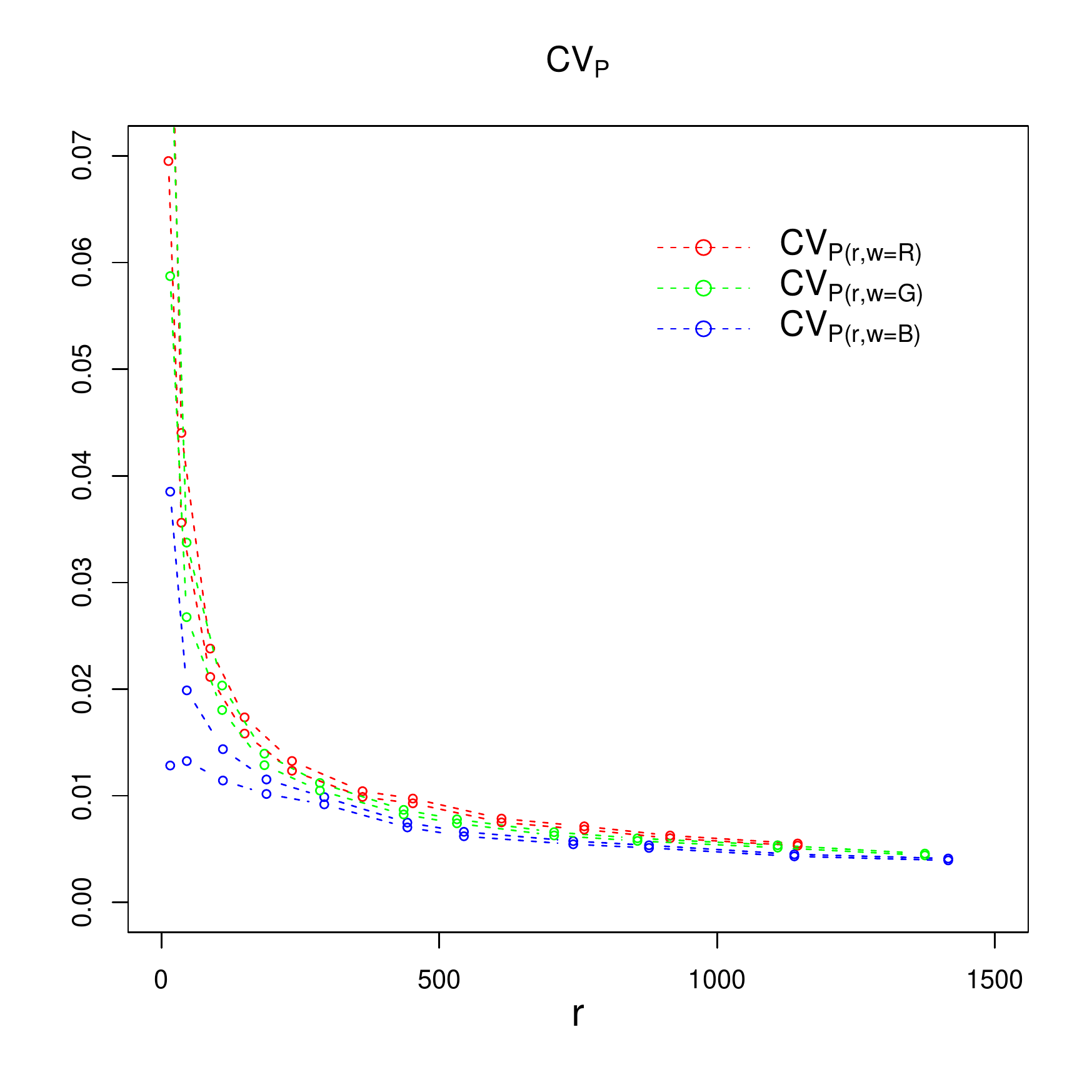} 
\end{tabular}
\vspace{-4mm}
\caption{2.5\% and 97.5\% credible intervals for the coefficient of variation of the parameter $S$ ($CV_S$) (a), parameter $T$ ($CV_T$) (b), parameter $P$ ($CV_P$) (c), versus mean output-reflectance $\mu_{\scriptscriptstyle 0}$ and wavelengths $w$.}
\label{ch3_fig04}
\end{figure*}

Figure \ref{ch3_fig04} shows the coefficients of variation of the parameters  $\bm{S}$ ($CV_S$), $\bm{T}$ ($CV_T$) and $\bm{P}$ ($CV_P$). The coefficient of variation $CV_S$, seen in Figure \ref{ch3_fig04}(a), is mainly constant with respect to reflectance, since the linear-multiplicative effect of the electrons $\bm{\mu}_e$ was removed, except for the lowest values of reflectance where $\bm{\sigma}_{\scriptscriptstyle S}$ has a non-linear behaviour (see Figure \ref{ch3_fig03}(a)). $CV_S$ represents the mean proportion of noise, relative to the input signal, caused by the variability of the normal variable $dK_i$ (or effect of interpixel differences when generating electrons (PRNU)), and was estimated around 0.004. Which means that the mean PRNU noise is of 0.4\% of the input signal, except at the lowest reflectances that reached up to 5\%.

The coefficient of variation $CV_T$ (Figure \ref{ch3_fig04}(b)) is constant since the linear effect of the light intensity on the current noise was removed. $CV_T$ represents the mean proportion of noise, relative to the input signal, caused by the free electrons generated by thermal effects in a pixel (current noise), and was estimated around 0.01. Which means that the mean current noise is of 1\% of the input signal.

As commented above, the slope of the noise effect $\sigma_{\scriptscriptstyle P}(r,w)$ of parameter $P_{it}(r,w)$ (Figure \ref{ch3_fig03}(d)) stems from photon noise (variance of the electrons $de_{it}(r,w)$; see equations (\ref{ch3:eq:electrons}), (\ref{ch3:eq:photon2}) and (\ref{ch3:eq:de})) that increases with the square root of the mean number of electrons. In fact, when computing the coefficient of variation $CV_P$ in Figure \ref{ch3_fig04}(c), it can be observed that the resulting slope is very similar to $1/sqrt(\mu_{\scriptscriptstyle K} \! \cdot \! \mu_{\scriptscriptstyle e}(r,w))$ which is the coefficient of variation of the photon noise in output of digital numbers. Thus, the mean proportion of noise, relative to the input signal, decline inversely proportional to the square root of reflectance, from over 0.03 at very low reflectances, to 0.004 at the highest reflectances. Which means that the mean photon noise range from over 3\% of the input signal at very low reflectances, to 0.4\% at the highest reflectances.

It can be stated, therefore, that the linear effect of the reflectance does not imply a lost of quality in the signal, since the coefficient of variation remains equal. However, it is an exception for the photon noise which does imply a lost of quality at low values of reflectance, as shown in its coefficient of variation in Figure \ref{ch3_fig04}(c). It is due to the inherent dependency of the variance of the electrons on the reflectance.

\section{Model checking and validation} \label{ch3:modelchecking}

In order to do posterior model checking against the observed data, an additional representative set $\mathfrak{D}$ of sample data for model testing which has not been taken part in fitting the model is available, as was stated in Section \ref{ch3:experiment}.

\begin{figure*}[t]
\centering
\begin{tabular}{ccc}
	{(a)} & {(b)} & {(c)} \\
  \includegraphics[trim = 0mm 15mm 8mm 20mm, clip, scale=0.33]{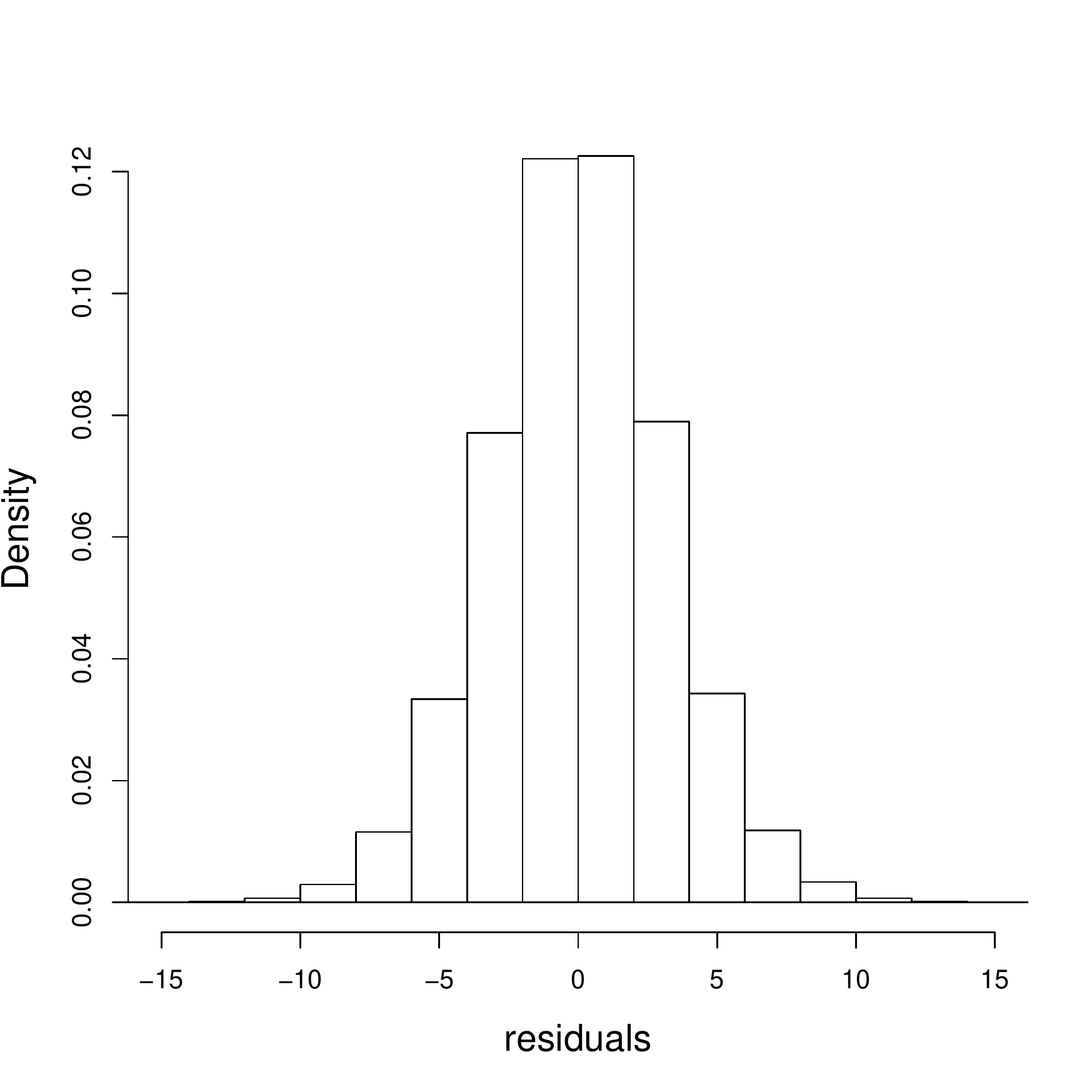} \hspace{-5mm}
  &\includegraphics[trim = 0mm 15mm 15mm 20mm, clip, scale=0.33]{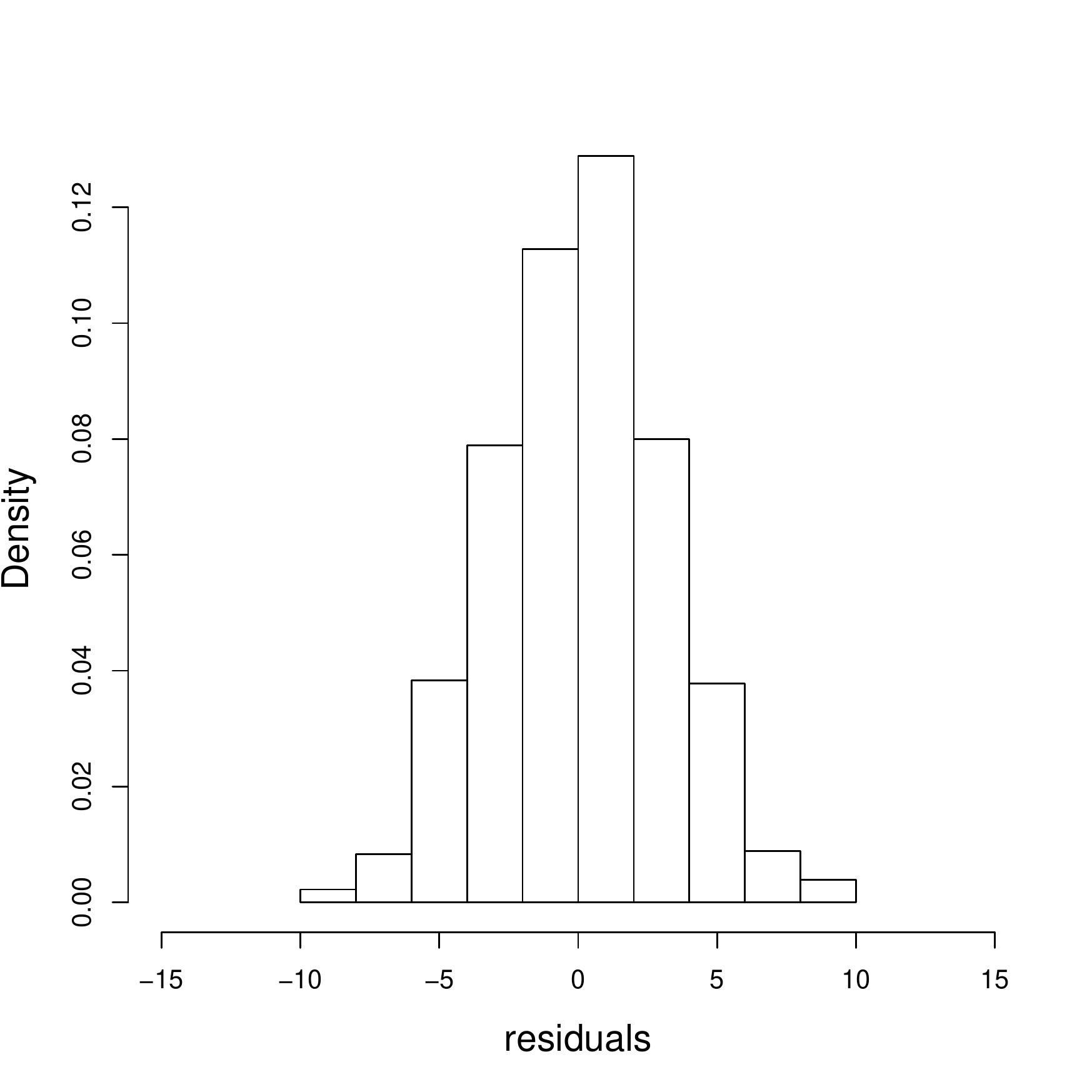}
  &\includegraphics[trim = 1mm 5mm 13mm 20mm, clip, scale=0.24]{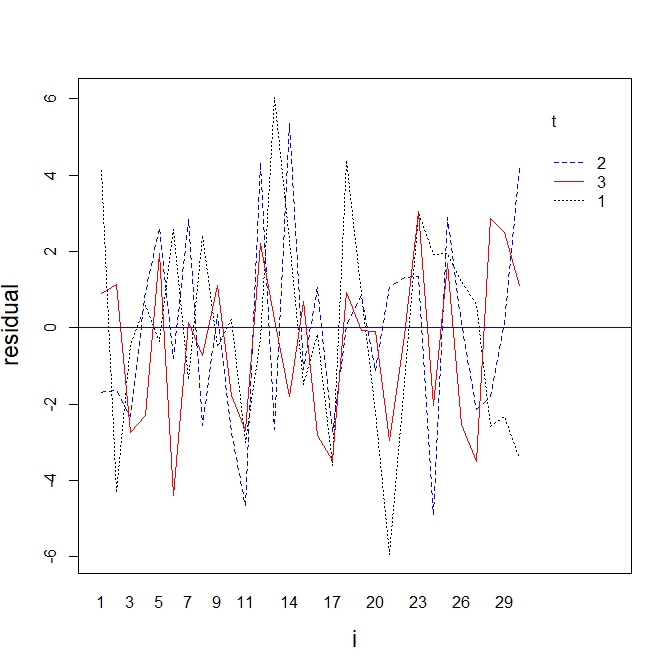}
\end{tabular}
  \caption{(a) Histogram of all residuals. (b) Histogram of residuals inside the group ($r$="1" and $w$=R). (c) Interaction plot between pixel (i) and exposure (t) dimensions inside group ($r$="1" and $w$=R). }
  \label{ch3_fig05}
\end{figure*}

First of all, common procedures for assessing normality and tendencies on the predictive residuals for this set $\mathfrak{D}$ of test data can be used. Figures \ref{ch3_fig05}(a) and \ref{ch3_fig05}(b) show histograms for all the predictive residuals and the predictive residuals inside the group ($r="1"$, $w=R$), respectively, which have the shape of a Gaussian distribution with zero mean. Figure \ref{ch3_fig05}(c) shows an interaction plot of the predictive residuals in order to check the independence between pixel (i) and exposure (t) dimensions. It is not noticed any kind of residual pixel pattern over time or any kind of residual temporal pattern over the pixel dimension. Despite a slight trending effect of the residuals with respect to reflectance, as seen in Figure \ref{ch3_fig03}(e), the residuals can be considered stable in practice as a function of reflectance, as stated in previous Section \ref{ch3:std}. Thus, it can be concluded that the residuals can be considered independent, random and normally distributed around zero, showing a good fit-to-the-data scenario.

\begin{figure}
\centering
\begin{tabular}{c}
   \includegraphics[trim = 0mm 5mm 15mm 20mm, clip, scale=0.25]{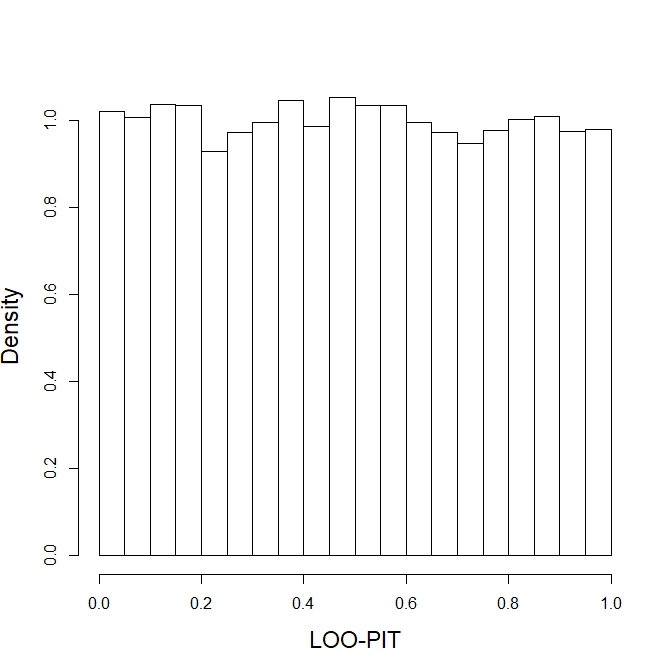}
\end{tabular}
  \caption{Histogram of the predictive posterior checks (LOO-PIT).}
  \label{ch3_fig06}
\end{figure}

Furthermore, the \textit{probability integral transformation} (PIT) is a rigorous statistic that can be used to assess whether the model predictive distributions are calibrated, that is, they are describing the model predictive uncertainty well. In case of good calibration of predictive distribution, PIT values are uniformly distributed \citep{Bayarri_2000}. They are based on computing the probability of a prediction $\tilde{y}_{it}$ to be lower or equal to its corresponding actual observation $y_{it}$ \citep{gelfand_1992,Gelman_2013}:
$$
\text{PIT}_{it} = P(\tilde{y}_{it} \leq y_{it}),
$$

\noindent where $i,t \in \mathfrak{D}$. Using sampling methods, computing the probability of a prediction being lower than or equal to the observed one is straightforward through the collection of simulated values for that prediction. The frequency histogram of PIT values showed in Figure \ref{ch3_fig06} is close to a uniform distribution, which means that the model predictions are well calibrated and points to a good and reliable approximation to the real process observed by the data.

Next, having assessed the calibration of predictive distributions, a global measure of model closeness to data needs to be computed, and it can be assessed using the \textit{root mean square predictive error} (RMSE). The RMSE evaluates, by averaging over all checking observations of the test dataset $\mathfrak{D}$, how far new data is from the model by using the distance (error) between the actual observation $y_{it}$ and the predictive mean $\tilde{y}_{it}$. And it results:
$$
\text{RMSE} = \sqrt{\frac{1}{|\mathfrak{D}|}\sum_{(i,t) \in \mathfrak{D}} (y_{it}  - \tilde{y}_{it})^2} = 3.55.
$$

\noindent where $|\mathfrak{D}|$ denotes the cardinality of $\mathfrak{D}$. Compared to the dynamic range of the experimentation (between 0 and 1500 digital numbers), a RMSE of 3.55 digital numbers shows that the model is accurate and close to data.

\section{Discussion} \label{ch3:discussion}
In the introduction, it has been stated that some of the advantages of Bayesian inference over classical estimation methods based on point estimates. 

The present work aims to argue and show the reliability and accuracy of Bayesian modeling and inference, by its ability to define proper prior probability distributions and to infer full posterior probability distributions for the parameters of interest \citep{Gelman_2013}. This differs from and contrasts with the fixed parameter definitions and point estimates of the classical methods \citep{browne2006comparison,raiko2006bayesian,bishop2006pattern}.

Furthermore, it has been emphasized that the inherent capability of propagating uncertainty among quantities of the Bayesian approach \citep{brown2014applied,gelman2006data,Gelman_2013}, in contrast to classical methods and, in particular, in contrast to the rigidness and the highly propagating error of nested independent point estimate computations of the commonly used Photon-transfer method for estimating sensor noise.

The Photon transfer method \citep{dierks2004sensitivity,international2003iso}, is considered as the standard for electronic noise characterization. However standard it is subject to some assumptions, such as:
\begin{itemize}
\item[-] Linear sensitivity (photo-response) of the sensor, i.e., the radiometric response (grey level values) increases linearly with the number of photons received. 
\item[-] All noise components are stationary and white with respect to time and space. The parameters describing the noise component are invariant with respect to time and space. 
\item[-] Only the total quantum efficiency is wavelength dependent, i.e., the effects caused by light of different wavelengths can be linearly superimposed. 
\end{itemize}

\noindent If these conditions are not fulfilled, the computed parameters by the Photon-transfer method are meaningless \citep{tsin2001statistical}. Photon-transfer method is based on the photo-response noise with and without light to determine all the parameters characterizing completely the sensor radiometry. The Photon-transfer method uses the property of spatial non-uniformities of a sensor array being the same for every exposure, to remove the effect of the spatial non-uniformity by differentiating two images. If temporal non-uniformity is present in the behaviour of a sensor, that is, the mean response is not stationary with respect to time, then the estimate does not represent the different photo-response among pixels. Therefore, the computed parameters by the Photon-transfer method are meaningless.

However, using statistical modeling it is not needed to make nested independent computations but estimating all the components at once in a model. Non-linear effects for the photo-response of a sensor can be easily considered in a statistical modeling approach, by using non-linear functions in classical methods and non-parametric models in a Bayesian approach. In our work, due to the fact that real reflectance values are unknown and only a few reflectance patches were measured, parameters $\bm{\mu}_0$ were defined as categorical factors that also allow for modeling non-linear effects. Otherwise, it could be defined, for example, a non-parametric prior distribution or a splines model for parameters $\bm{\mu}_0$.

The present work is a novel attempt to model the sensor noise parameters. For this reason, the data-generating model with the defining effects has been formulated as found in the general state of the art literature, where noise parameters $\bm{S}$ (PRNU), $\bm{F}$ (FPN) and $\bm{T}$ (current noise) are considered completely random-structured effects and stationary with respect to space, for the parameters $\bm{S}$ and $\bm{F}$, and time, for the parameter $\bm{T}$. However, correlated effects in space and time can be naturally considered by using the Bayesian approach. For example, there may be some imaging sensors that show systematic FPN patterns, instead of being completely random as the one considered here. Nevertheless, handling this issue is straightforward in the Bayesian approach provided that appropriate prior distributions with correlated effects are defined. Furthermore, non-stationary noise parameters, such as spatial effects varying in time or time effects varying in space or space-time effects, might also be feasibly considered in a Bayesian framework.

These powerful and flexible modeling features of Bayesian hierarchical models \citep{dai2017predicting,coley2017bayesian} are promising in image sensing, opening the door to formulate new data-generating models where new effects could be investigated.

\section{Conclusion}\label{ch3:conclu}

The formulation of a Bayesian hierarchical model with different and independent random effects allowed us to identify the major noise components that take place in image sensing. The approach presented in this work provided a useful interpretation and an accurate estimation of the image sensing noise parameters. Bayesian modeling permitted a reliable definition of parameters as probability distributions and accurately propagated uncertainty among quantities in a fully probabilistic model.

Our focus has been on the analysis and interpretation of parameters $\bm{\sigma}_{\scriptscriptstyle S}$, $\sigma_{\scriptscriptstyle F}$, $\bm{\sigma}_{\scriptscriptstyle T}$, $\bm{\sigma}_{\scriptscriptstyle P}$ and $\sigma$ which represent the mean noise of the parameters PRNU ($\bm{\sigma}_{\scriptscriptstyle S}$), FPN ($\sigma_{\scriptscriptstyle F}$), current noise ($\bm{\sigma}_{\scriptscriptstyle T}$), photon noise ($\bm{\sigma}_{\scriptscriptstyle P}$), and amplifier, flicker and quantization noises ($\sigma$).

Furthermore, the dependency of the estimated noise parameters $\bm{\sigma}_{\scriptscriptstyle P}$, $\bm{\sigma}_{\scriptscriptstyle T}$ and $\bm{\sigma}_{\scriptscriptstyle S}$ on reflectance suggested the computation of the coefficients of variation (noise and mean intensity ratio) in order to remove the linear effect of the mean level of reflectance imaged. Thus, they can be considered useful quantities to be compared among sensors as a discriminatory power of the signal. The coefficients of variation of the noise are larger at lower reflectances than higher reflectances due to the effect of the photon noise. The photon noise effects decline inversely proportional to the square root of reflectance, from approximately 3\% of the registered signal at very low reflectances, to 0.4\% at the highest reflectances examined in the experimentation. Which reveals that high image intensity values are preferred to lower image intensity values for applications such as, for example, image pattern recognition tasks. On the other hand, the effects of the current noise and PRNU are practically constant of around 1\% and 0.4\% of the registered signal, respectively.

A brief comparison of our approach with the existing standards has been made. The assumptions of linearity and stationary of the parameters, considered by existing standards, can be easily overcome by statistical modeling and, especially, using a Bayesian approach. 

The Bayesian multilevel random-effects modeling approach presented in this paper is a general methodology that can be applied to any other imaging sensor or camera, under different experimental, independently of the dynamic ranges. Future research will be focused on assessing the assumptions under complete randomness in time (current noise) and space (PRNU and FPN). For this purpose, appropriate prior distributions with correlated effects have to be considered. Furthermore, all noise parameters must be modeled using their exactly defining probability distributions, instead of approximating them by Normal distributions, which is an usual assumption in image sensing.

\section*{Acknowledgment}

The authors gratefully acknowledge the support from the Instituto de Salud Carlos III and FEDER, project PI18/00881, as well as grants PPIC-2014-001 and SBPLY/17/180501/000491/1  funded by Consejer\'ia de Educaci\'on, Cultura y Deportes (JCCM) and FEDER.

\bibliography{references}

\end{document}